\documentclass[graphics, twocolumn, usenatbib]{mn2e}
\usepackage{times}
\usepackage{natbib} 
\usepackage{epsfig} 
\usepackage{graphicx} 
\usepackage{color}
\usepackage{aas_macros} 
\usepackage{amssymb}
\usepackage{amsmath}
\usepackage[title]{appendix}
\usepackage{hyperref}	% Hyperlinks
\hypersetup{colorlinks=true,linkcolor=blue,citecolor=blue,filecolor=blue,urlcolor=blue}
\usepackage[caption=false]{subfig}

\newcommand{\enzo}{\texttt{Enzo~}}
\newcommand{\enzoc}{\texttt{Enzo}}

\newcommand{\kms} {km $\rm{s^{-1}}$}
\newcommand{\mpch} {\rm $h^{-1}$ Mpc\,\,} 
\newcommand{\kpch} {\rm $h^{-1}$ kpc\,\,} 
\newcommand{\msolar} {$\rm{M_{\odot}}~$}
\newcommand{\msolarc} {$\rm{M_{\odot}}$}

\newcommand{\molH} {$\rm{H_2}$~}
\newcommand{\molHc} {$\rm{H_2}$}

\newcommand{\smartstar} {\texttt{SmartStar~}}

\newcommand{\smartstarc} {\texttt{SmartStar}}

\begin{document}
\title[]{Super-Eddington Accretion and Feedback from the First Massive Seed Black Holes}

\author[J.A. Regan, T.P. Downes, M. Volonteri, R. S. Beckmann, A. Lupi, M. Trebitsch \& Y. Dubois] 
       {John A. Regan$^{1}$\thanks{E-mail:john.regan@dcu.ie, Marie Sk\l odowska-Curie Fellow},
         Turlough P. Downes$^{1}$, Marta Volonteri$^{2}$, Ricarda Beckmann$^{2}$,\newauthor
         Alessandro Lupi$^{2}$, Maxime Trebitsch$^{2}$ \& Yohan Dubois$^{2}$\\
         $^1$Centre for Astrophysics \& Relativity, School of Mathematical Sciences,
         Dublin City University, Glasnevin, D09 W6Y4, Ireland\\
         $^2$Sorbonne Universites, UPMC Univ Paris 6 et CNRS, UMR 7095, Institut
         d’Astrophysique de Paris, 98 bis bd Arago, 75014 Paris, France
}
%Start things off
%\date{Accepted XXX. Received YYY; in original form ZZZ}
\pubyear{2018}
\label{firstpage}
\pagerange{\pageref{firstpage}--\pageref{lastpage}}

%Make the Title
\maketitle

 %%%%%%%%%%%%%%%%%%%%%%%%%%%%%%%%%%%%%%%%%%%%%%%%%%%
%Abstract time
\begin{abstract}
  Super-Eddington accretion onto massive black hole seeds may be commonplace in
  the early Universe, where the conditions exist for rapid accretion.
  Direct collapse black holes are often invoked as a possible solution to the observation of
  super massive black holes (SMBHs) in the pre-reionisation Universe.
  We investigate here how feedback, mainly in the form of bipolar jets, from
  super-Eddington accreting seed black holes will affect their subsequent growth.
  We find that, nearly independent of the mass loading of the bipolar jets, the violent
  outflows generated by the jets evacuate a region of approximately 0.1 pc surrounding the black hole
  seed. However, the jet outflows are unable to break free of the halo and their impact is limited
  to the immediate vicinity of the black hole. The outflows suppress any accretion
  for approximately a dynamical time. The gas then cools,
  recombines and falls back to the centre where high accretion rates are again observed. The
  overall effect is to create an effective accretion rate with values of between 0.1 and 0.5 times
  the Eddington rate. If this episodic accretion rate is maintained for order 500 million years then
  the black hole will increase in mass by a factor of between 3 and 300 but far short of the factor of
  $10^4$ required for the seeds to become the SMBHs observed at $z>6$.
  Therefore, direct collapse
  black holes born into atomic cooling haloes and which experience strong negative mechanical
  feedback will require external influences (e.g. rapid major mergers with other haloes) to
  promote efficient accretion and reach SMBH masses within a few hundred million years.
\end{abstract}

%%%%%%%%%%%%%%%%%%%%%%%%%%%%%%%%%%%%%%%%%%%%%%%%%%%
%keywords time
%%%%%%%%%%%%%%%%%%%%%%%%%%%%%%%%%%%%%%%%%%%%%%%%%%%
%Introduction time

%%%%%%%%%%%%%%%%%%%%%%%%%%%%%%%%%%%%%%%%%%%%%%%%%%%%%%%%%%%%%%%%%%%%%%%%%%%%

\section{Introduction} \label{Sec:Introduction}

The discovery of super-massive black holes (SMBHs) with masses in excess of $10^9$ \msolar
at redshifts greater than $z = 6$~\citep{Fan_06, Mortlock_11, Tang_2019} presents a significant difficulty for
theories of black hole formation and growth. Black holes are expected to form as the end point of
massive stars. Black holes forming from the first generation of massive Population III (PopIII) stars
have initial seed masses close to their final stellar mass \citep[e.g.]{Woosley_2002}.
However, these PopIII remnant black holes are expected to be born ``starving'' \citep{Whalen_2004,
  OShea_2005b, Wang_2006, Johnson_2007, Milosavljevic_2009, Alvarez_2009, Jeon_2012}.
A more recent study by
\cite{Smith_2018} using a sample of approximately 15,000 PopIII remnant black holes from the
Renaissance simulation suite \citep{Xu_2013, Chen_2014, OShea_2015} saw no evidence
for significant accretion onto the remnant black holes with PopIII remnants increasing their mass
by at most 10\% over several hundred million years\footnote{It should be noted that this study
  investigated the accretion onto the black holes in post-processing only and neglected the impact of
  dynamical friction which may have increased the accretion rates}.
The black holes are typically born into
low density environments due to an initial supernova explosion which results in severely
stunted growth. For PopIII stars within the direct collapse window \citep{Heger_2003} the
black hole initially experiences rapid accretion, however, the phase is short lived, with high density
gas quickly consumed by further star formation. Even if PopIII remnant stars can remain in a region
of high density, where local star formation is suppressed, a PopIII remnant would need to accrete
at the Eddington limit for several hundred megayears in order to reach a mass of close to a billion
solar masses by a redshift of 6. Such a scenario is very unlikely based on current research. \\
\indent In light of this, several other pathways have been explored to attempt to understand
the appearance of SMBHs in the first billion years of the Universe. The scenarios have broadly been
divided into light seed scenarios and heavy seed scenarios. Light seed scenarios encompass
mechanisms where the initial black hole mass is ``light'' (M$_{\rm{init}} \sim$ 100 \msolarc) but grows
rapidly. The PopIII remnant case falls under the light seed scenarios, as do cases where
initially light seeds rapidly merge together to form a more massive object. Several authors have
considered a scenario where stellar collisions in high-redshift,
dense star clusters lead to the runaway growth of a single star
\citep{PortegiesZwart_2004, Gurkan_2004, Gurkan_2006, Freitag_2006, Omukai_2008,
  Devecchi_2008,Katz_2015, Habouzit_2017}.
In this scenario, a dense stellar cluster
becomes unstable to gravitational collapse leading to the merger of a significant number of the stars
in the cluster and the formation of a single massive star through mass segregation.
The most massive stars which emerge from the cluster are expected to have initial masses of the
order of 1000 \msolarc. 
\indent Alternatively, there are a number of scenarios where a heavy seed (M$_{\rm{init}} \gtrsim 10^4$
\msolarc) may emerge. In the centre of rapidly accreting atomic cooling haloes, which are metal-free,
a supermassive star (SMS) is expected to form
\citep{Eisenstein_1995b, Bromm_2003, Regan_2009b, Regan_2009}. SMS formation requires very
high accretion rates in excess of 0.01 \msolarc/yr
\citep{Begelman_2006,Begelman_2008,Schleicher_2013, Sakurai_2016} to inflate the envelope
around the protostar and sustain
a super-massive (or possibly a quasi) star \citep{Hosokawa_2012, Hosokawa_2013, Inayoshi_2014,
  Umeda_2016, Woods_2017, Haemmerle_2017, Haemmerle_2017b}.
If accretion rates in excess of 0.01 \msolarc/yr can be sustained for the
lifetime of the star then the star is expected to collapse into a massive black hole seed at the
end of its lifetime, either through the General Relativistic instability \citep{Chandrasekhar_1964b}
or after the star runs out of nuclear fuel. The final mass of the SMS is expected to be well in
excess of $10^4$ \msolarc. The collapse into a direct collapse black hole then leaves a black hole seed
with a large initial mass. If no supernova explosion occurs then the black hole can be born into
a region with a plentiful supply of gas from which it can accrete. \\
\indent Typically, the environmental
conditions required for the heavy seed model require strong sources of nearby Lyman-Werner radiation,
which can efficiently dissociate \molH \citep{Dijkstra_2008, Dijkstra_2014, Visbal_2014b, Regan_2017}.
However, dynamical processes which collisionally dissociate \molH may also induce the
correct environmental conditions for direct collapse black holes
\citep{Mayer_2010, Inayoshi_2012, Fernandez_2014,Mayer_2014,Inayoshi_2015}.
Similarly, relative streaming velocities between baryons and
dark matter following recombination \citep{Tseliakhovich_2010} has been investigated by several
authors in the context of the first massive black holes \citep{Tanaka_2014, Schauer_2017, Hirano_2017},
with promising results. In summary, several pathways remain open to generating environmental conditions
for the formation of massive back hole seeds. \\
\indent Accretion onto the black hole in either scenario will determine the future growth of the
black hole. The Eddington accretion rate can be derived by equating the gravitational force of a
black hole to the radiative force experienced by the in-falling matter. The resulting force balance
applies in the case of a spherically symmetric collapse, with the Eddington
accretion rate given by
\begin{equation}
  \dot{M}_{\rm Edd} = {{4 \pi G M_{\rm BH} m_p} \over {\eta \sigma_T c}}
\end{equation}
where $M_{\rm BH}$ is the black hole mass, $m_p$ is the proton mass, $\eta$ is the radiative efficiency,
$\sigma_T$ is the Thomson scattering cross section and $c$ is the speed of light. However,
it is known that in non-spherically symmetric circumstances, the Eddington rate can be breached and
super-Eddington accretion may persist. In this case, accretion can then proceed extremely rapidly.
Numerous models of super-Eddington accretion exist. For example, the slim disk model of
super-Eddington accretion was originally developed by \cite{Abramowicz_1988} to investigate
scenarios where the Eddington limitation could be broken. Super-Eddington accretion models of
accretion onto stellar mass black holes have recently been investigated by a number of
authors \citep{Sadowski_2009, Sadowski_2014, Sadowski_2016, Sadowski_2016a, Jiang_2017} with results
consistently showing that super-Eddington accretion can be achieved with observational evidence
also mounting to support super-Eddington accretion \citep[e.g.][]{Du_2018}.\\
\indent Super-Eddington accretion has been shown, through numerical models, to generate
powerful bipolar jets, which become active as the accretion rate exceeds the Eddington rate. These
jets, though highly collimated, have the potential to shut off the very accretion flow that is
driving the jets, and regulate the accretion flow to values sub-Eddington. Previous investigations
have included only radiative feedback from BH seeds accreting at super-Eddington rates
\citep{Pacucci_2015a,Lupi_2016,Sakurai_2016a, Pezzulli_2016, Pezzulli_2017,
  Sugimura_2017, Toyouchi_2018, Inayoshi_2018}. 
Furthermore, the works listed above have been, by necessity, somewhat idealised.
  We investigate here a self-consistent 3-D cosmological setting where an embryonic black hole
  seed finds itself at the centre of a strong accretion flow. We
investigate if an initial seed mass black holes accreting  above the Eddington rate can sustain a
large accretion in the presence of bipolar jets.

\section{Numerical Framework} \label{Sec:Model}
\noindent In this study we have used the publicly available adaptive mesh refinement code
\enzoc\footnote{http://enzo-project.org/} to study the birth of a massive black hole seed from a
SMS. We have utilised the \smartstar particles introduced in \cite{Regan_2018a} and
augmented them with subgrid prescriptions specific to a black hole seed as we now discuss. 

\subsection{Enzo}
\enzoc\footnote{Changeset:48882af312bc} \citep{Enzo_2014} is an adaptive mesh refinement code
ideally suited to simulations of the high redshift universe. Gravity in \enzo is solved using
a fast Fourier technique \citep{Hockney_1988}, which solves the Poisson equation on the root grid
at each timestep. On subgrids, the boundary
conditions are interpolated to the subgrids and the Poisson equation is then solved at each timestep.
Dark matter is represented using particles, with each particle stored on the highest refinement grid
available to it and thus the particle has the same timestep as the gas on that grid. The
particle densities are interpolated using the cloud-in-cell technique
onto the grid and solved at the same time as the gas potential.
\enzo contains several hydrodynamics schemes to solve the Euler equation. To model the physics of
jet launching, we use the Zeus hydrodynamic solver \citep{Stone_1992, Stone_1992b}.
A known limitation of the Zeus solver is the inclusion of artificial viscosity that can cause
spurious heating of gas upstream from a shock front \citep{Anninos_1994}. However, the correct
Rankine-Hugoniot jump conditions are nonetheless achieved. The very high resolution of our
simulations and in particular the small number of cells over which jets are launched, goes some
way towards mitigating these effects. Furthermore, the Zeus solver is very robust and able to
follow the sharp discontinuities that arise as the jets are launched. Zeus is second order
accurate in space and first order accurate in time. \\
\indent Chemistry is an important component in following the collapse of (ideal) gas. We use the
\texttt{Grackle}\footnote{https://grackle.readthedocs.org/}$^,$\footnote{Changeset:482876c71f73}
\citep{Grackle} library to follow the evolution of ten individual species:
${\rm H}, {\rm H}^+, {\rm He}, {\rm He}^+,  {\rm He}^{++}, {\rm e}^-,$ 
$\rm{H_2}, \rm{H_2^+}\, \rm{H^-} \rm{and}\ \rm{HeH^+}$. We adopt here the 26 reaction network
determined by \cite{Glover_2015a} as the most appropriate network for solving the chemical
equations required by gas of primordial composition with no metal pollution and exposed to an external
radiation source. The network includes the most 
up-to-date rates as described in \cite{GloverJappsen_2007},  \cite{GloverAbel_2008},
\cite{GloverSavin_2009}, \cite{Coppola_2011}, \cite{Coppola_2012},  \cite{Glover_2015a},
\cite{Glover_2015b},  \cite{Latif_2015}. The cooling mechanisms
included in the model are collisional excitation cooling, collisional ionisation cooling,
recombination cooling, bremsstrahlung and Compton cooling off the cosmic microwave background.
  \molH line cooling is explicitly followed as part of the Grackle chemistry network following
  the prescription given by \cite{Abel_1997}.

%%%%%%%%%%%%%%%%%FIGURE 1%%%%%%%%%%%%%%%%%%%%%%%%%%%%%%%%%%%%%%%%%%%%%%%
\begin{figure*}
  \centering 
  \begin{minipage}{175mm}      \begin{center}
      \centerline{
        \includegraphics[width=0.47\textwidth,height=6.5cm]{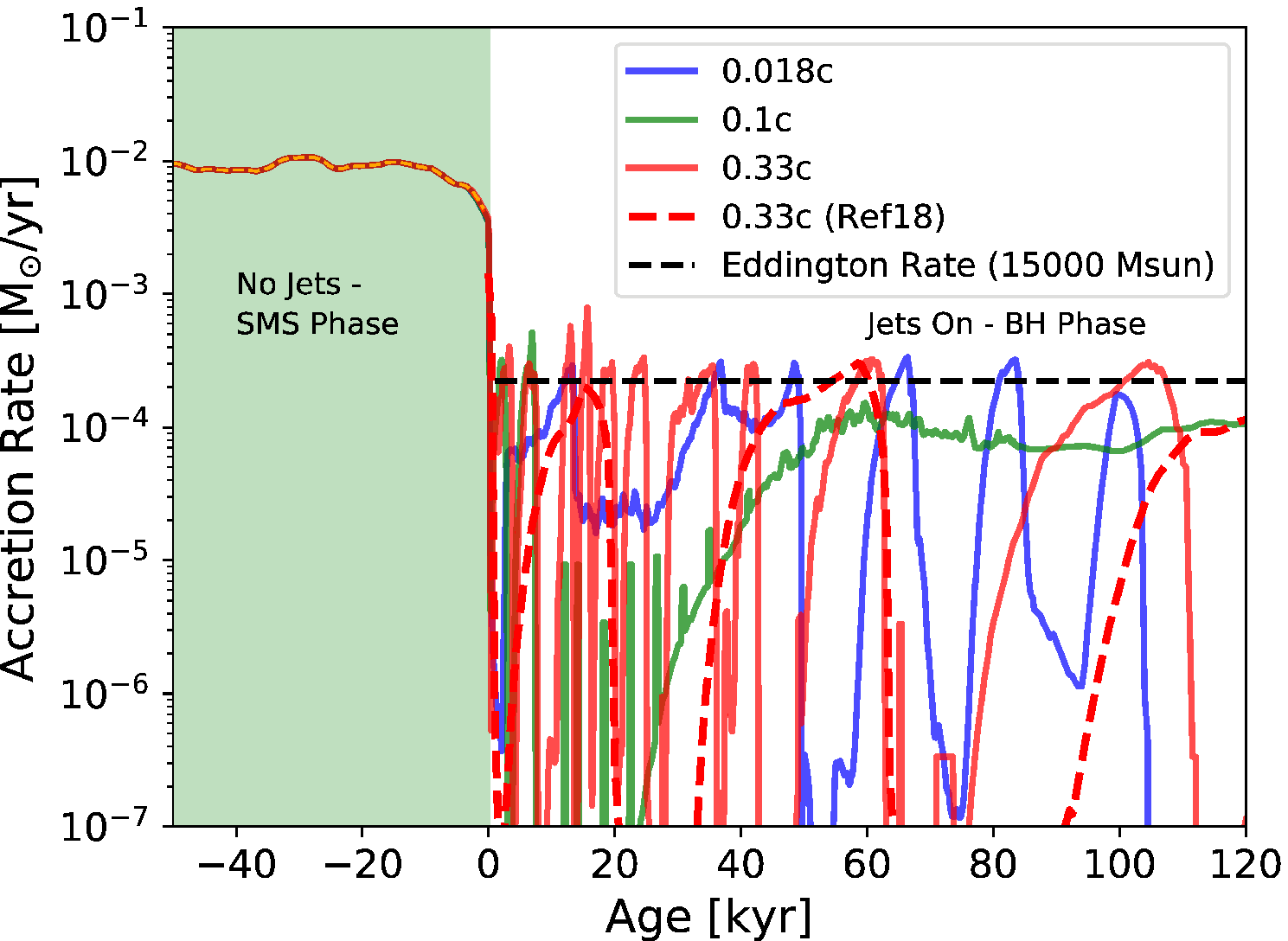}
        \includegraphics[width=0.47\textwidth,height=6.5cm]{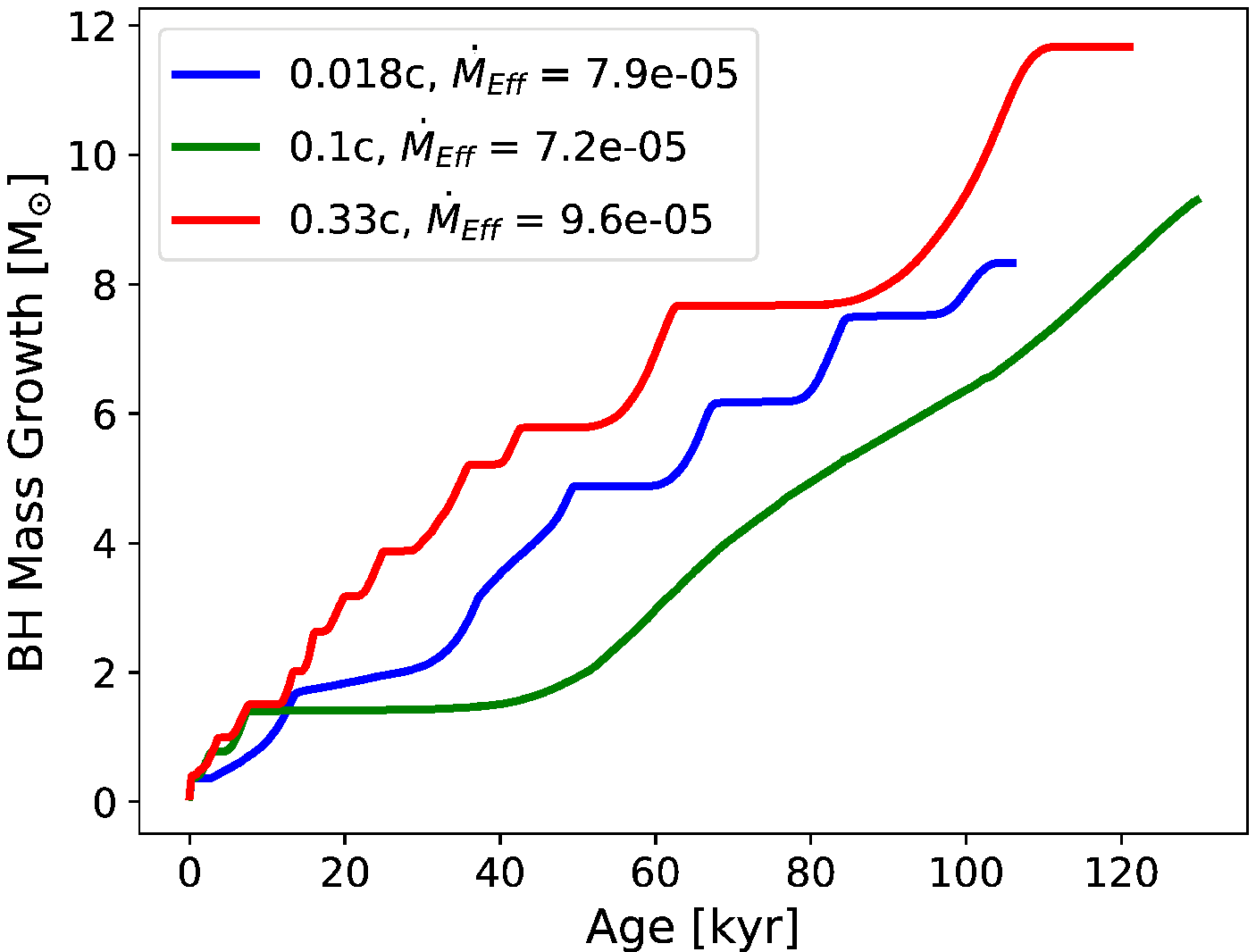}}
      \caption[]
      {\label{MassAccretionRate}
        \textit{Left Panel}: Mass accretion rates onto the black hole for different mass loading prescriptions.
        The x-axis runs from -50 kyr to 120 kyr. The black hole forms at T = 0 years. The negative
          timescale represents the SMS phase of the object. In this phase these is only non-ionising
          radiative feedback and no jets. Jets turn on once the object collapses to a BH at T = 0. The
          appearance of the jets signals a severe shift in the accretion history of the object. 
        All simulations show that the jets effectively limit accretion to below the Eddington
        rate. Once accretion exceeds the Eddington limit the jets turn on disrupting the
        accretion flow. Differences between the impact of different prescriptions are
        clearly visible but similarities exist. The dynamical time for the highest density
        gas in the centre of the halo is a few thousand years and so gas falls back to the
        centre on this timescale. This is evident from the spikes for each prescription. 
          The dashed red line is a lower resolution simulation using jets with launch speeds of 0.33c.
          Qualitatively the behaviour is similar to the higher resolution runs. 
        \textit{Right Panel}: The mass growth of each seed black hole. The initial mass of each seed is identical,
        M$_{seed} = 15904$ \msolarc. The effective accretion rate onto each seed is calculated over 100,000 years and
        found to be almost $\dot{M} \sim 10^{-4}$ \msolarc/yr in each case. This results in a mass increase of roughly 10 \msolar
        in each case over the first 100,000 of the seed black holes existence. 
    }
  \end{center} \end{minipage}
\end{figure*}

%%%%%%%%%%%%%%%%%%%%%%%%%%%%%%%%%%%%%%%%%%%%%%%%%%%%%%%%%%%%%%%%%%%%%%%%%%%%

\subsection{Simulation Setup}
\noindent The simulation explored here is the ``Ref20\_100J21\_OT'' simulation from
\cite{Regan_2018b}, hereafter R18b.
This simulation used a Lyman-Werner background of 100 J$_{21}$ to dissociate \molH and allowed for the
formation of an atomic cooling halo in which a SMS can form. The SMS formed at a redshift of
$z = 24.7$. The maximum resolution of the simulation was set to $2.5 \times 10^{-4} \rm{pc} (\sim 50$ au).
At this resolution, resolving the outer envelope of the SMS becomes possible. In R18b the
simulation was run for 250 kyr at which point a single SMS was accreting at approximately
$10^{-2}$ \msolarc/yr (see Figure 2 from R18b) and had achieved a mass of approximately
15,000 \msolarc. We begin the simulation for this study from this point. We briefly review
the original simulation for completeness. 

The original simulation was  run within a cosmological box of 2 \mpch (comoving), 
  on a root grid of $256^3$ and with three levels of nested grids. The grid nesting and
  initial conditions were created using MUSIC \citep{Hahn_2011}. Within the most refined region
  (i.e. level 3) the dark matter particle mass is $\sim$ 103 \msolarc. In order to further increase
  the dark matter resolution of our simulations, we split the dark matter particles according to the
  prescription of \cite{Kitsionas_2002}, as described in \cite{Regan_2015}. We split particles
  centered on the position of the final collapse as found from lower resolution simulations within a
  region with a comoving side length of 43.75 h$^{-1}$ kpc. Each particle is split into 13 daughter
  particles resulting in a final high resolution region with a dark matter particle mass of
  $\sim$ 8 \msolarc. The particle splitting is done at a redshift of 40, well before the collapse of
  the target halo. Convergence testing to study the impact of lower dark matter particle masses was
  discussed in \cite{Regan_2015}. \\
  \indent The baryon resolution is set by the size of the grid cells. In the highest resolution region 
  this corresponds to approximately 0.48  \kpch comoving (before adaptive refinement). Setting the
  maximum refinement level for this simulation to 20 results in a maximum resolution of
  $2.5 \times 10^{-4}$ pc. 
  As is standard in simulations of this type, refinement is triggered in \enzo  when certain 
  user defined thresholds are exceeded. The refinement criteria used in this work were based on
  three physical measurements: (1) The dark matter particle over-density, (2) the baryon
  over-density and (3) the Jeans length. The first two criteria introduce additional meshes
  when the over-density of a grid cell with respect to the mean gas or dark matter density exceeds
  8.0. Furthermore, we set the \emph{MinimumMassForRefinementExponent} parameter
  to $-0.1$ making the refinement more aggressive for the baryon and dark matter overdensity
  and hence making the behaviour of the adaptive mesh ``super-Lagrangian'' in nature
  (see \cite{Enzo_2014} for further details). This technique also reduces the threshold for
  refinement as higher densities are reached. For the final criteria we set the number of cells
  per Jeans length to be 32 in these runs. \\
  \indent In order to suppress PopIII star formation and allow the
  simulation to form pristine atomic cooling haloes, we imposed an artificial Lyman-Werner background.
  We set the effective temperature of the background radiation field to
  T$\rm{_{eff}} = 30000$ K. This background temperature suitably models the spectrum of a
  population of young stars \citep{WolcottGreen_2012, Sugimura_2014, Latif_2015}.
  The effective temperature of the background is important as the radiation temperature
  determines the dominant photo-dissociation reaction set in the irradiated halo. This in turn
  leads to a value of J$_{crit}$ - the flux above which complete isothermal collapse of the
  irradiated halo is observed due to the complete suppression of \molHc.\\
  \indent 
As the gas density increases in high density regions, hydro codes, including \enzoc, require a method
to convert the high density gas into stars in many cases. This is done to deal
with gas which has reached the  maximum allowed refinement level of the simulation and for which
further collapse is being artificially suppressed through artificial pressure support.
Within  the ``Ref20\_100J21\_OT'', simulation particles were introduced once the following
criteria were met:
\begin{enumerate}
\item The cell is at the highest refinement level
\item The cell exceeds the Jeans density
\item The flow around the cell is converging along all axes
\item The cooling time of the cell is less than the freefall time
\item The cell is at a local minimum of the gravitational potential
\end{enumerate}

As described in R18, all ``stars'' which form are initially  assumed to be stars with low surface
temperatures that are appropriate for main sequence SMSs and less massive proto-stars on the
Hayashi track. As long as the accretion rate remains above a critical value of
$\dot{M_*} \gtrsim 0.04$ \msolarc/yr \citep{Sakurai_2016}, the star remains a SMS. If the accretion
rate drops below this critical value, the star contracts and becomes a PopIII star. In 
``Ref20\_100J21\_OT'', the accretion rate dropped below the critical value shortly after
formation, after approximately 25 kyr. Nonetheless, the accretion rate remained high even
though the ionising radiation
from the PopIII was able to ionise and heat some of the gas immediately surrounding the proto-star.
Similar results were observed in the simulations of \cite{Chon_2017b}.
The accretion rate remained relatively constant at around $\dot{M_*} \sim 0.01$ \msolarc/yr for the
duration of the simulation ($\sim 250$ kyr). At this point we now allow the massive PopIII star to 
transition to a massive black hole seed. Ideally, we would have allowed the PopIII to continue to
accrete until it either ran out of nuclear fuel (after $\sim 10^6$ years) or reached the
GR instability (after reaching a mass of M$_{SMS} \sim 5 \times 10^5$ \msolarc). However, the
computational expense in running the simulation at this refinement level is extreme and the physics
of massive PopIII star evolution is insufficiently understood to pursue this course. Instead we,
prematurely, convert the star particle into a black hole particle in order to
study the impact that this change will have on the surrounding material and the
accretion onto the black hole.

\subsection{Accretion onto the black hole} \label{accretion}
\indent The accretion onto the black hole particle is similar to the accretion mechanism used
to accrete onto the star particle. The particle can accrete gas within its accretion radius (4 cells)
and it can merge with other \smartstar particles. Accretion onto the \smartstar is determined by
calculating the flux of gas across the accretion surface.
\begin{equation}
  \dot{M} = 4\pi \int_S {\rho v_r^- r^2 dr}
\end{equation}
where $\dot{M}$ is the mass accretion rate, $S$ is the surface over which we integrate, $\rho$ is the
density of the cells intersecting the surface, $v_r^-$ is the velocity of cells intersecting
the surface and which have negative radial velocities and $r$ is our surface's radius.
The surface, $S$, is
the surface of a sphere with radius the accretion radius. As noted above we set the accretion
radius to be 4 cells, we choose to fix this radius independently of the resolution or the mass of
the \smartstarc. We do this so as to be as accurate as possible when calculating the accretion
rate, any mass travelling radially inward at a distance of four cells from the \smartstar is taken
to be accreted onto the \smartstar - we therefore strive for the maximum possible physical
resolution. \\
\indent As an alternative to directly measuring accretion using the mass flux method described above
we can also calculate the accretion rate on the black hole using the Bondi-Hoyle
prescription \citep{HoyleLyttleton_1939, HoyleLyttleton_1940, HoyleLyttleton_1940a, Bondi_1952}.
As described in \cite{Krumholz_2004} we use the following approximate formula which was originally
given in this approximate form by \cite{Ruffert_1994} and \cite{Ruffert_1994a}
\begin{equation}
  \dot{M} = 4 \pi \rho_{\infty} r_{BH}^2 ( \lambda^2 c_{\infty}^2 + v_{\infty}^2)^{1/2}
\end{equation}
where $\rho_{\infty}$ is the density of gas at the Bondi-Hoyle radius,  $r_{BH}$ is the Bondi-Hoyle
radius, $c_{\infty}$ is the sound speed at infinity (in the host cell in this case)
and $v_{\infty}$ is also the relative velocity of the sink particle
and the gas in the host cell of the black hole. $\lambda$ is a constant of order unity,
we follow \cite{Krumholz_2004} in that regard and use $\lambda = e^{3/2}/4 \sim 1.120$
  throughout. While,
at the high resolution we are able to evolve our simulations at, the mass flux approach is more
accurate we use the Bondi-Hoyle prescription immediately after jets are launched. We do this to
prevent the procedure from calculating spurious accretion rates due to the large mechanical feedback
from the jets. After 50 further timesteps the accretion procedure automatically reverts to the mass
flux method. We determined this number (50) after careful testing of the mass flux
  accretion rate against the Bondi-Hoyle accretion rate. We did not employ the Bondi-Hoyle
  accretion rate for the entire time because we found that during testing (using Enzo and Ramses)
  that the flux accretion method performed significantly better in several analytic tests
  particularly the \cite{Shu_1977} collapse test. 
\\
\indent The spatial extend of our most refined cells is dx $\sim 2.5 \times 10^{-4}$ pc
  ($\sim$ 50 au). The accretion radius is therefore $R_{acc} \sim 10^{-3}$ pc. For a 15,000 \msolar
  black hole surrounded by a gaseous medium at approximately 10,000 K the Bondi-Hoyle radius is
  approximately
  $r_{BH} \gtrsim 10^{-2}$ pc and therefore we are resolving the Bondi-Hoyle radius extremely well
  at that point. However, as the feedback from the accretion leads to bi-polar jets the medium can
  heat up to close to $10^6$ K at the edge of the accretion zone leading to a $r_{BH} \lesssim
  10^{-5}$ pc. We are now no longer resolving the Bondi-Radius and hence we apply the kernel
  weighting techniques advocated by \cite{Krumholz_2004}.\\
\indent In the scenario where the characteristic scale (in this case the Bondi-Hoyle scale) is
significantly below the resolution of the simulation it would be erroneous to set the accretion
rate derived from a scale significantly beyond the true accretion scale. In this case we apply a
kernel weighted averaging procedure to the accretion rate calculated numerically. In doing this
we follow equations 13 and 14 from \cite{Krumholz_2004}\\
%\begin{align}
%  {$\Delta$} x  / {4} %& \ \ \ \ \ r$_{BH}$ $lt$ %{{\Delta} x \over {4}}, \\
%  r_K =  r$_{BH}$ & \ \ \ \ \ %{{\Delta} x \over {4}} $\le$ r$_{BH}$ $\le$  {{r$_{acc}$} \over {2}}, \\
 % {{r_{acc} \over {2}} & \ \ \ \ \ %r_{BH} > {{r_{acc}} \over {2}}, \\  
%\end{align}
\begin{equation}
r_{K} =
\begin{cases}
  {dx  \over {4}} &  \text{if \ } r_{BH} < {dx \over {4}}, \\
  r_{BH} & \text{if \ } {dx \over {4}} \le r_{BH} \le  {{r_{acc}} \over {2}}, \\
  {{r_{acc}} \over {2}} & \text{if \ } r_{BH} > {{r_{acc}} \over {2}} \\
\end{cases}
\end{equation}
where $r_{K}$ is the kernel radius and $r_{acc}$ is the accretion radius (4 cells). For all cells then
within the accretion radius the kernel weight, $w$, is applied according to
\begin{equation}
  w \propto \text{exp}(-r^2/r_{K}^2)
\end{equation}
where the normalisation is calculated by computing the sum of the weights. \\
\indent The accretion onto the star is calculated at each timestep,
however this is likely to be a very noisy metric. To alleviate this to some degree we average
the accretion rate over hundreds of timesteps typically corresponding to between 10 and 100 years.
The average accretion rate is then used as the actual accretion rate.
The accretion rate is added as an attribute to each star and hence a full accretion history
of every \smartstar is outputted as part of every snapshot. \\

\subsection{Feedback from the black hole} \label{feedback}
The feedback from an accreting black hole is primarily determined by the radiative efficiency of the
disk, $\eta_{disk}$. $\eta_{disk}$ is typically set to a value close to 0.1 for a non-rotating black
hole. For these simulation we use a value very close to this, $\eta_{disk} = 0.103$, which we derive
by explicitly accounting for the spin of the black hole \citep[e.g.][]{Sadowski_2016a}
\begin{equation}
  \eta_{disk} = 1 - \sqrt(1 -\frac{2.0}{3.0 R_{\rm ISCO}(a)})
\end{equation}

where a is the spin parameter of the black hole which we set to $a = 0.7$ and ${R_{\rm ISCO}}$ is
a parameterisation of the inner most stable orbit given by \cite{Abramowicz_2013}
\begin{equation}
  R_{\rm ISCO} = R_G * \Big( 3 + Z_2 - [(3 - Z_1)*(3 + Z_1 + 2Z_2)]^{\frac{1}{2}} \Big)
\end{equation}
where $Z_1 = 1 + (1-a^2)^{1/3}\Big((1+a)^{1/3}+(1-a)^{1/3}\Big)$, \\
$Z_2 = \Big(3*a^2+Z_1^2\Big)^{1/2}$ and
$R_G$ is the gravitational radius, $R_G = GM/c^2$. The accretion rate onto the black hole must now
be modified to account for the energy that is returned to the outer medium from the accretion
\begin{equation}
  \dot{M}_{BH} = \dot{M} * (1 - \eta_{disk})
\end{equation}
where $ \dot{M}_{BH}$ is now the mass accretion rate onto the black hole while $\dot{M}$ is the
numerically determined accretion rate onto the black hole as described in \S \ref{accretion}.
The feedback from the black hole can now be further decomposed into radiative feedback from the
disk and mechanical feedback from a jet component.

\subsubsection{Radiative Feedback}
To model the radiative feedback from the black hole 
we assume a multi-colour disk for the accretion disk and then a fit a corona with a
power law \citep[e.g.][]{Done_2012}. We divide the energy radiated equally between the multi-colour
disk and the power law component.
The radiative feedback within \enzo is modelled using the ray tracing module
MORAY \citep{WiseAbel_2011}, which discretises the radiation into a set of finite energy bins which
are then transported outwards from the black hole particle. We split the radiation into five
energy bins from infrared up to hard X-rays. The energy bins used are 2.0 eV, 12.8 eV, 19.1eV,
217.3 eV and 5190 eV with the actual value of the luminosity at each timestep determined by the
accretion rate at that timestep. The fractional energy in each energy bin is then determined by the
accretion rate onto the black hole and the mass of the black hole.
For super-Eddington adjustments to the radiative feedback
we employ the fit from \cite{Madau_2014} who themselves use \cite{Sadowski_2009} to derive the
fit. In this case the luminosity is calculated as

\begin{equation}
  \frac{L}{L_E} = A(a) \Big( \frac{0.985}{\dot{M}_E/\dot{M}_{BH} +B(a)} + \frac{0.015}{\dot{M_E}/\dot{M}_{BH} +C(a)} \Big)
\end{equation}
where the functions A, B, C scale with the spin of the black hole, $a$, as
\begin{align}
  A(a) &= (0.9663 - 0.9292a)^{-0.5639} \\
  B(a) &= (4.627 - 4.445a)^{-0.5524} \\
  C(a) &= (827.3 - 718.1a)^{-0.7060}
\end{align}
and $\dot{M_E}$ is the Eddington mass accretion rate. The luminosity per solar mass is adjusted
in this case compared to the thin disk model but the energy bins and energy fraction per bin
remain unchanged. In essence the radiative efficiency is reduced, as expected. \\
\indent For the cases considered here we limit the radiative feedback component to non-ionising
radiation only - i.e. we use the first two energy bins of our model only. We do this for
three reasons. Firstly, the simulations are computationally expensive
and in order to reduce the computational expense we limit  the radiative feedback to
being optically thin and below the ionisation potential of hydrogen. Secondly,  the
fraction of energy emitted as ionising radiation falls off rapidly as the accretion rate decreases and
by considering only the infrared and Lyman-Werner components we are nonetheless still capturing
the bulk of the radiative processes. Finally, in this work we are primarily interested in
investigating the impact of mechanical feedback (i.e. jets) on the ability of seed black holes
to accrete effectively and hence neglecting the ionising radiation component allows us to do that.
A full treatment of the radiative feedback will be considered in an upcoming study. Appendix
  A contains a detailed discussion of the procedure used in determining the spectral energy
  distribution (SED) of the black hole simulated in this paper. 

%%%%%%%%%%%%%%%%%FIGURE 2%%%%%%%%%%%%%%%%%%%%%%%%%%%%%%%%%%%%%%%%%%%%%%%
\begin{figure*}
  \centering 
  \begin{minipage}{175mm}      \begin{center}
      \centerline{
        \includegraphics[width=9cm]{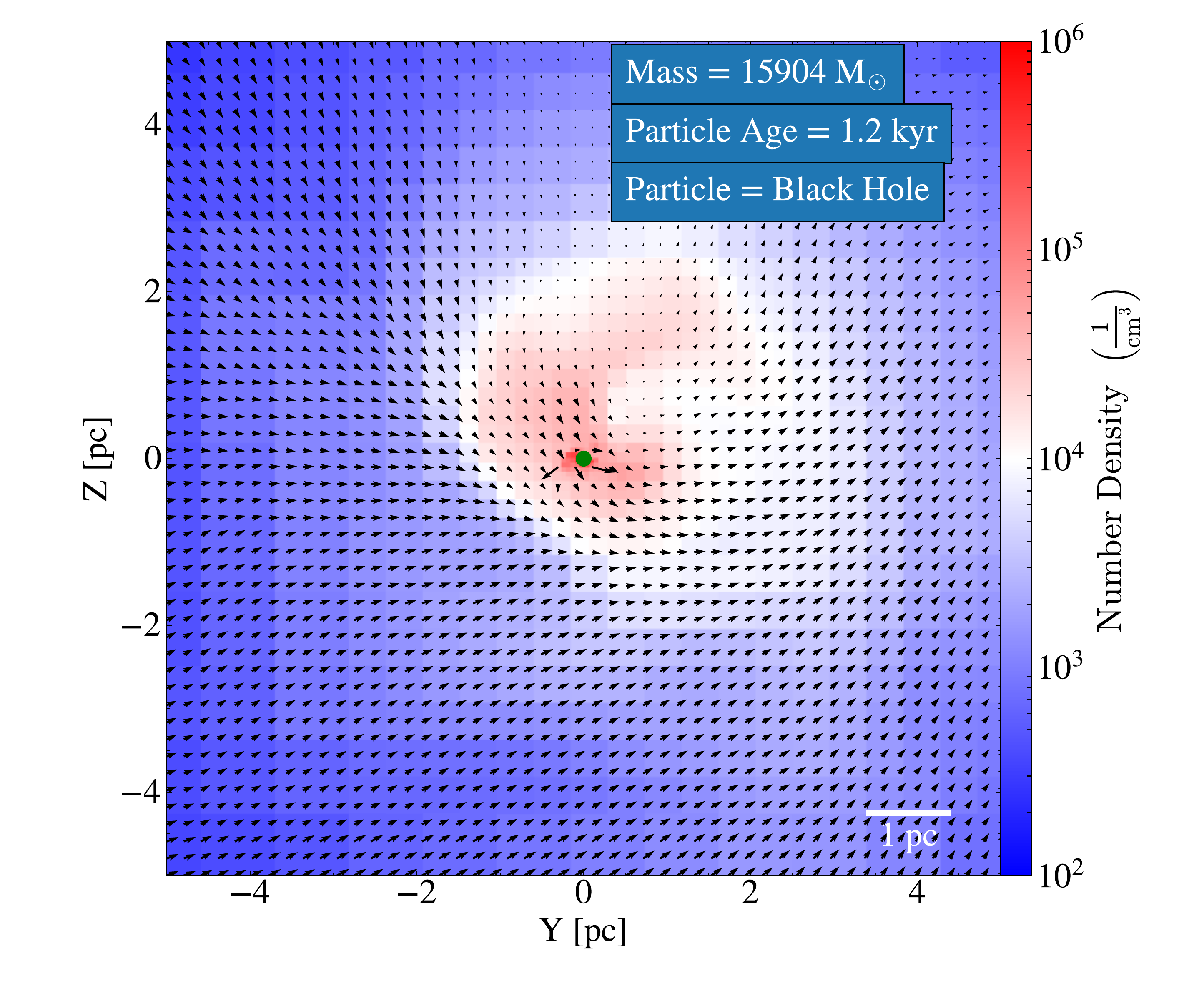}
        \includegraphics[width=9cm]{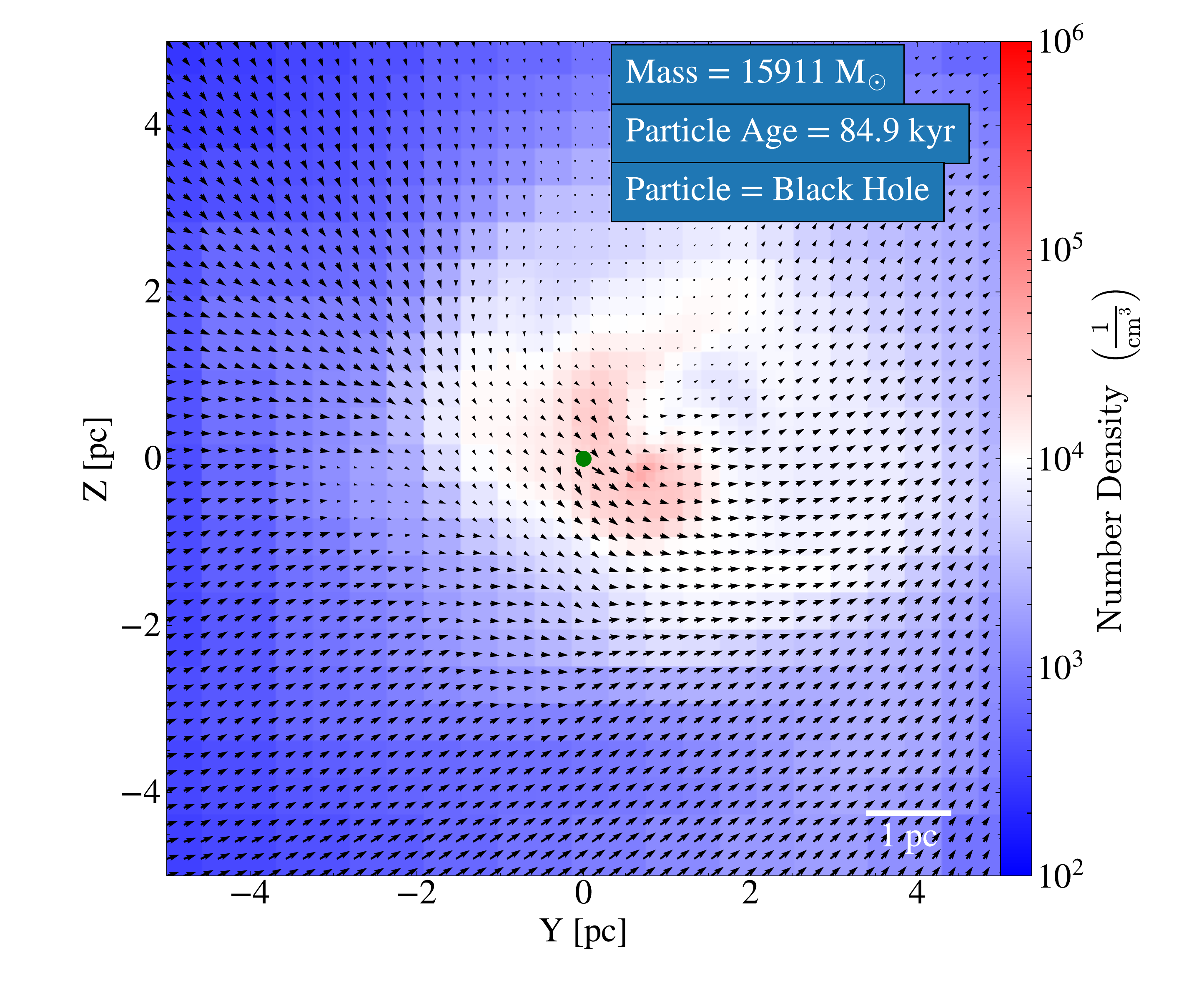}}
        \caption[]
        {\label{Projection6000}
          The number density of the gas in a 10 pc volume surrounding the black hole
          (marked in green). The
          velocity of jets in this projection is 6000 \kms. The velocity of the jets
          is marked with arrows to give the direction. As the accretion rate exceeds
          the Eddington rate bipolar jets are launched from around the black hole (e.g.
          in the left hand panel). We make the visualisations near the
          start of the black hole evolution and near the end. In the left hand panel the
          black hole has just released an outflow - which can be seen as the two longer than
          average velocity arrows. However, at this scale of a few parsecs the impact of the
          jets is very mild. The jets have a strong local effect, as we will see, but globally
          they have little effect on the halo. The impact of the jets is clearly visible in
          Figure \ref{FourPanel} where we zoom in on the region surrounding the black hole.
          Note, that the
          radial extent of the black hole has been greatly exaggerated for this plot.
        }
      \end{center} \end{minipage}
  \end{figure*}

%%%%%%%%%%%%%%%%%%%%%%%%%%%%%%%%%%%%%%%%%%%%%%%%%%%%%%%%%%%%%%%%%%%%%%%%%%%%

 %%%%%%%%%%%%%%%%%FIGURE 3%%%%%%%%%%%%%%%%%%%%%%%%%%%%%%%%%%%%%%%%%%%%%%%
\begin{figure*}
  \centering 
  \begin{minipage}{175mm}      \begin{center}
      \centerline{
        \includegraphics[width=18cm]{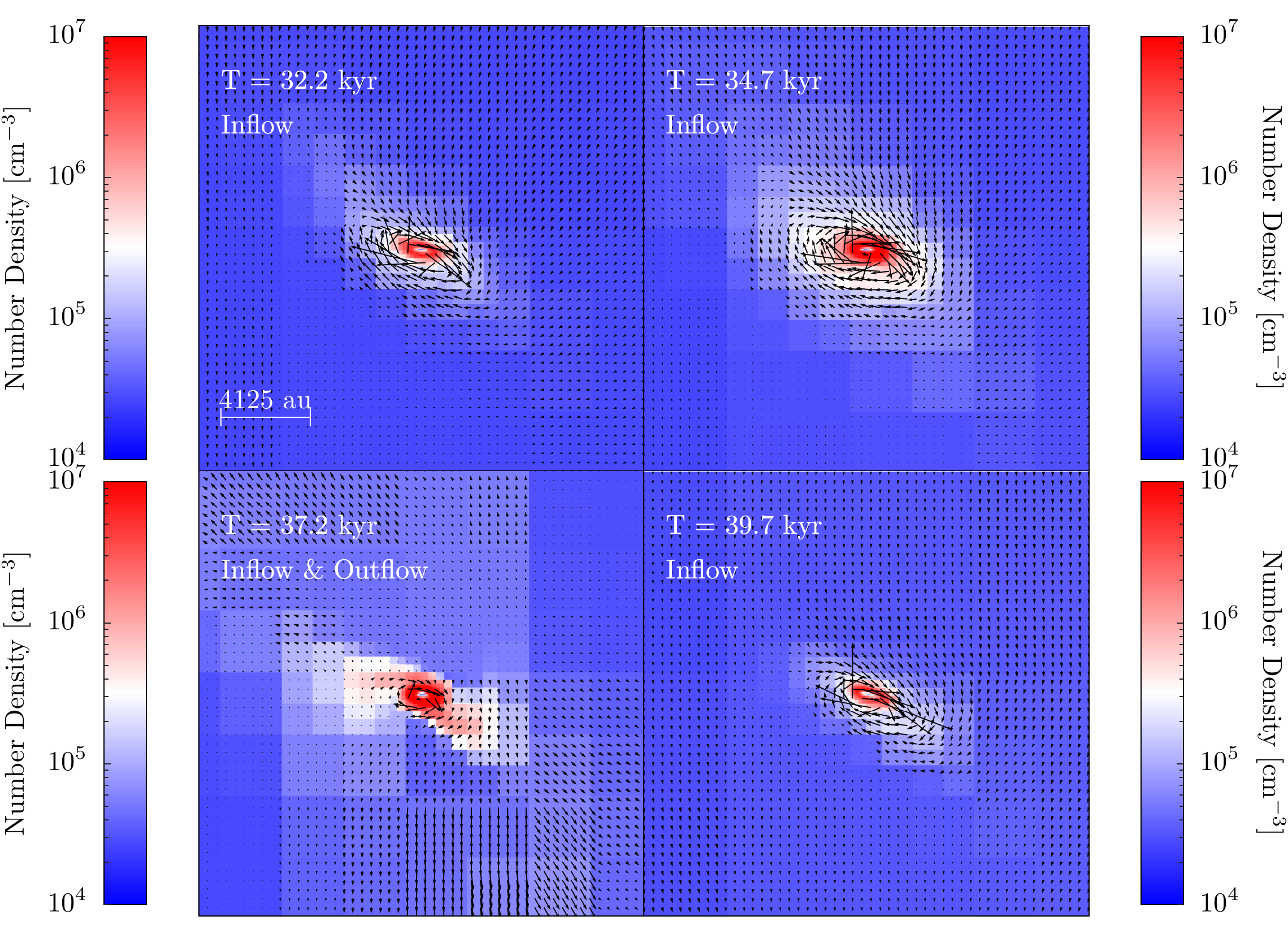}}
        \caption[]
                {\label{FourPanel}
                  A ``zoom-in'' projection onto a period of accretion (inflow) followed
                  by a jet event (outflow). The projections are made for the simulation
                  with jet outflows of 6,000 \kms between approximately 30 kyr and 40 kyr after
                  the formation of the black hole. The scale line shown in the top left panel
                  gives a scale of 4125 au. Each panel covers a region of 0.1 pc on the side.
                  A super-Eddington outflow occurs approximately
                  at $T \sim 37$ kyr (see also Figure \ref{MassAccretionRate}) following
                  accretion above the Eddington
                  limit. Further accretion is then able to continue following the outflow. 
        }
      \end{center} \end{minipage}
  \end{figure*}

%%%%%%%%%%%%%%%%%%%%%%%%%%%%%%%%%%%%%%%%%%%%%%%%%%%%%%%%%%%%%%%%%%%%%%%%%%%%

%%%%%%%%%%%%%%%%%FIGURE 4%%%%%%%%%%%%%%%%%%%%%%%%%%%%%%%%%%%%%%%%%%%%%%%
\begin{figure*}
  \centering 
  \begin{minipage}{175mm}      \begin{center}
      \centerline{
        \includegraphics[width=18cm]{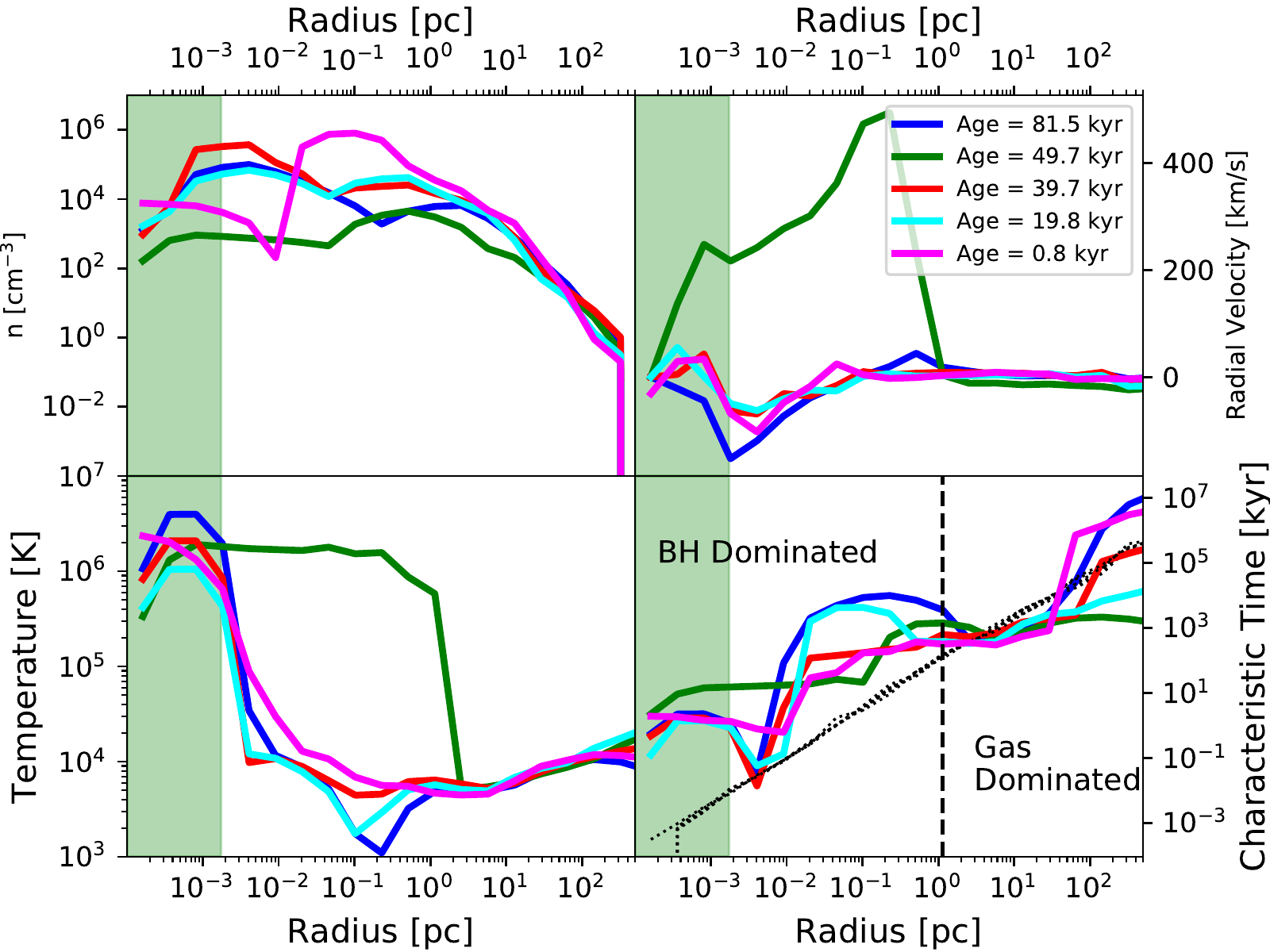}}
        \caption[]
                {\label{Radial6000}
                  Ray profile plots of the number density, temperature, electron
                  fraction and the 'characteristic' time of the gas at times between
                  seed black hole formation and the end of the
                  simulation after 100,000 years. The characteristic time plot contains
                    both the
                  dynamical time (black dotted lines) and the cooling time of the gas (coloured
                  lines). The green shaded region on the extreme left of each panel is the accretion
                  zone of the SmartStar particle (i.e. the black hole). Values inside this region
                  should be treated with caution as the gas is accreted from within this region and
                  we advise readers not to draw conclusions from values inside the accretion zone.
                  The profiles are 1D profiles, perpendicular to the
                  angular momentum vector (i.e. through the accretion disk). The impact of the jets is
                  most strikingly seen in the green line in this figure (compare the
                  time to Figure \ref{MassAccretionRate}). At age $\sim$ 49.7 kyr
                  the accretion rate exceeds the Eddington rate driving a jet and
                  mass outflow at 6,000 \kms. The temperature rapidly increases to over
                  $10^6$ K out to almost 1 pc and the gas is strongly ionised.
                  The flow of the gas is clearly seen in the radial velocity profile plot (top right).
                  Within the disk the gas is accreting onto the disk and black hole but further out
                  the outflow, as a result of the jet, is clearly visible. 
                  The dynamical time of the gas increases linearly with radius with the
                    cooling
                  timescales always longer than the dynamical timescales. At radii
                  less that approximately 1 pc the dynamical timescale is dominated by the black
                  hole.
                  Outside of that time the gas is dominated by the self gravity of the
                  infalling gas. The dynamical time for the gas within 1 pc is less than 10 kyr
                  and so the gas is able fall back to the centre on approximately
                  this timescale. 
        }
      \end{center} \end{minipage}
  \end{figure*}

%%%%%%%%%%%%%%%%%%%%%%%%%%%%%%%%%%%%%%%%%%%%%%%%%%%%%%%%%%%%%%%%%%%%%%%%%%%%
\subsubsection{Mechanical Feedback} \label{MechanicalFeedback}
Microphysical models of the physics of accretion disks have shown that bipolar jets produced
predominantly by the tangling of magnetic field lines are a robust feature of super-Eddington
accretion. Jets appear to also be present at low accretion rates, most frequently when the accretion rates falls below $10^{-3}$ M$_{Edd}$ \citep[e.g.][and references therein]{Merloni_2008,Sadowski_2016a},
but in this paper we want to focus on the effects of jets launched during super-Eddington phases, therefore we do not initiate jets for accretion rates below the Eddington rate and instead all of
the feedback is radiative in that case (below the ionisation threshold of hydrogen as discussed above).

To calculate how much energy is mechanical output in the super-Eddington regime, 
we again follow the models of \cite[][equations 42 through 46]{Sadowski_2016a}. In this case the total jet luminosity
is given by
\begin{equation}
  L_{jet} = \eta_{jet} \dot{M}_{\rm BH} c^2
\end{equation}
where $\eta_{jet}$ is the jet efficiency factor given by
\citep{Sadowski_2016a}
\begin{equation}
   \eta_{jet} = 1.3 a^2
\end{equation}
This efficiency assumes maximum efficiency of the jet, where we have assumed a ``MAD''
  value of 1, making this an upper limit to the jet efficiency \citep{Sadowski_2016a}.
An additional complication in modelling jets is that jets are an inherently relativistic phenomenon and
their launch speed is close to the speed of light. Furthermore, the jets are launched on scales close to
R$_G$ which is far below the resolution of our simulations. 
Therefore, modelling both the speed and the initial launch radius of the jet are beyond the
capabilities of \enzoc. Hence we ``mass load'' the jet \citep[e.g][]{Ciotti_2001, Dubois_2012}
by adding additional mass to the
jet and by reducing the speed of the jet. This accounts for the assumption that the speed of the
jet will diminish as the jet entrains mass on its way from the black hole. The mass loading factor,
$\beta_{jet}$, 
is defined as
\begin{equation}
  \beta_{jet} = {\dot{M}_{jet} \over \dot{M}_{BH}}
\end{equation}
where $\dot{M}_{jet}$ is the amount of material ejected by the jet per unit time and again
$\dot{M}_{BH}$ is the mass projected to accrete onto the black hole surface. We now define the
jet 'kinetic' power, KE$_{jet}$, as in \cite{Kim_2011}
and equate it to the luminosity of the jet using conservation of energy to write
\begin{align}
  KE_{jet} &= \frac{1}{2}  \dot{M}_{jet} v_{jet}^2 \\
  KE_{jet}  &=  L_{jet}
\end{align}
we can then equate the mass loading factor, $\beta_{jet}$, with the velocity of the jet, $v_{jet}$,
and write
\begin{equation}
  \beta_{jet} = 2 \eta_{jet} \frac{c^2}{v_{jet}^2} \label{massloading}
\end{equation}
and we see that the mass loading of the jet and the velocity of the jet are degenerate as expected.
Typically, in numerical simulations $v_{jet}$ is set to be much less than c. For example setting
$v_{jet}$ = 0.1c gives $\beta_{jet} = 127.4$ while $v_{jet}$ = 0.01c gives $\beta_{jet} = 12740$. In both
cases $\beta_{jet} \gg 1$.\\
\indent Attempting to mass-load the jet by factors of up to $10^5$ can be problematic as there may
not be enough mass in the surrounding cells to do so. In this, rather common,
case we adjust the mass accretion rate onto the black hole so that the black hole can effectively
only accrete for a fraction of the current timestep. The fraction is calculated so there is sufficient
mass to load the jet. Ideally, we would like to decrease the timestep of the simulation such that the total mass required for a single accretion + feedback episode,
i.e. $M_{tot} = \Delta {M}_{BH} + \Delta M_{jet}  = (1 + \beta_{jet}) * M_{BH}$ is less than the
total mass available in the accretion region i.e. $< M_{acc}$ where
$\Delta {M}_{BH} = \dot{M}_{BH} \Delta t$. 
Practically, this will make the timestep unaffordably short, particularly in the super-Eddington
accretion regime, where both the amount of mass removed from the grid, and the total energy to be
returned to the grid, will be high. Instead we decrease the accretion time in this subgrid manner.
We do this by introducing a factor, $\epsilon_t$, which operates on the accretion rate modifying
both the actual accretion rate found for the black hole and the resulting jet ejection rate.
$\epsilon_t$ is calculated as
\begin{equation}
 \epsilon_t = \rm{min}(1.0, \frac{\dot{M}_{acc}}{\dot{M}} \frac{1.0}{1 - \eta_{disk}} \frac{1.0}{1 + \beta_{jet}})
\end{equation}
where $\dot{M}_{acc}$ is the maximum possible accretion rate, i.e. the total mass in the
accretion sphere divided by the timestep.
The above equation ensures mass conservation within the subgrid algorithm with $\epsilon_t$ fixed to
be always less than one. A further consequence of this approach is that the ejected mass is
very similar in all cases independent of the speed of the jet. Consider the following:
\begin{align}
  \dot{M}_{jet} &= \beta_{jet} (1 - \eta_{disk}) \dot{M} \frac{\dot{M}_{acc}}{\dot{M}} \frac{1.0}{1 - \eta_{disk}} \frac{1.0}{1 + \beta_{jet}} \\
  \dot{M}_{jet} &=  {\dot{M}_{acc} \over {\frac{1}{\beta_{jet}}} + 1}\\
  \dot{M}_{jet} &\sim \dot{M}_{acc} \ \ \ \ \ \ \ \ \ \ \ \rm{for\ \beta_{jet} \gg 1} \label{jetmass}
\end{align}
hence we have found that the mass ejected by the jet will be close to, but always less than,
the mass in the surrounding accretion sphere. This is expected since we need $\epsilon_t$ to be such
that there is always sufficient mass available to mass load the jet. Clearly, this is not
  as good as allowing the timestep to drop to the required value but recall that $\epsilon_t$ is only
  less than unity in the case where no mass is available for accretion and feedback at that timestep.
  It is therefore a practical approach to a resolution limited problem. In practice we find that for
  the vast majority of the time $\epsilon_t = 1$. \\
\indent Now that the algorithm for determining the mass of the jet and the speed of the jet has been
determined, it remains to describe how the jets are launched within the simulation. In this regard
we follow both \cite{Kim_2011} and \cite{Dubois_2012}. \cite{Kim_2011} use ``supercells''
within the \enzo grid hierarchy to launch the jets, effectively adding mass and velocity to cells
on the outer edge of a cone to launch the jet. As recommended by \cite{Dubois_2012} we insert
the jet at the maximum resolution and over the minimum number of cells as possible so as to have
the jet as collimated as possible. Typical jets are observed to be less than 1000 R$_G$ in radius
\citep[e.g.][]{Doeleman_2012} and so well below the resolution of our simulations - hence
we insert the jet over a limited number of cells (26). These 26 cells are the cells which are immediately
surrounding the black hole (i.e. $3^3$ - 1 = 26). (see also figure 2 from \cite{Kim_2011} for a pictorial
representation of the ``supercells''. Note that \cite{Kim_2011} injected velocity into cells
which were further from the black hole, in neighbour of neighbour cells and so over 98 cells ($5^3 - 2^3 = 98$)).
We inject cells close to the black hole with velocity, directed along the angular momentum vector and anti-parallel
to it resulting in bipolar jets which are as highly collimated as our resolution allows. The effect of this is to make the jet denser
relative to spreading the jet over a larger number of cells - the typical density of the jets launched
in our simulations is $\rho_{jet} \sim 10^{10}$ cm$^{-3}$.

%%%%%%%%%%%%%%%%%FIGURE 5%%%%%%%%%%%%%%%%%%%%%%%%%%%%%%%%%%%%%%%%%%%%%%%
\begin{figure*}
  \centering 
  \begin{minipage}{175mm}      \begin{center}
      \centerline{
        \includegraphics[width=18cm]{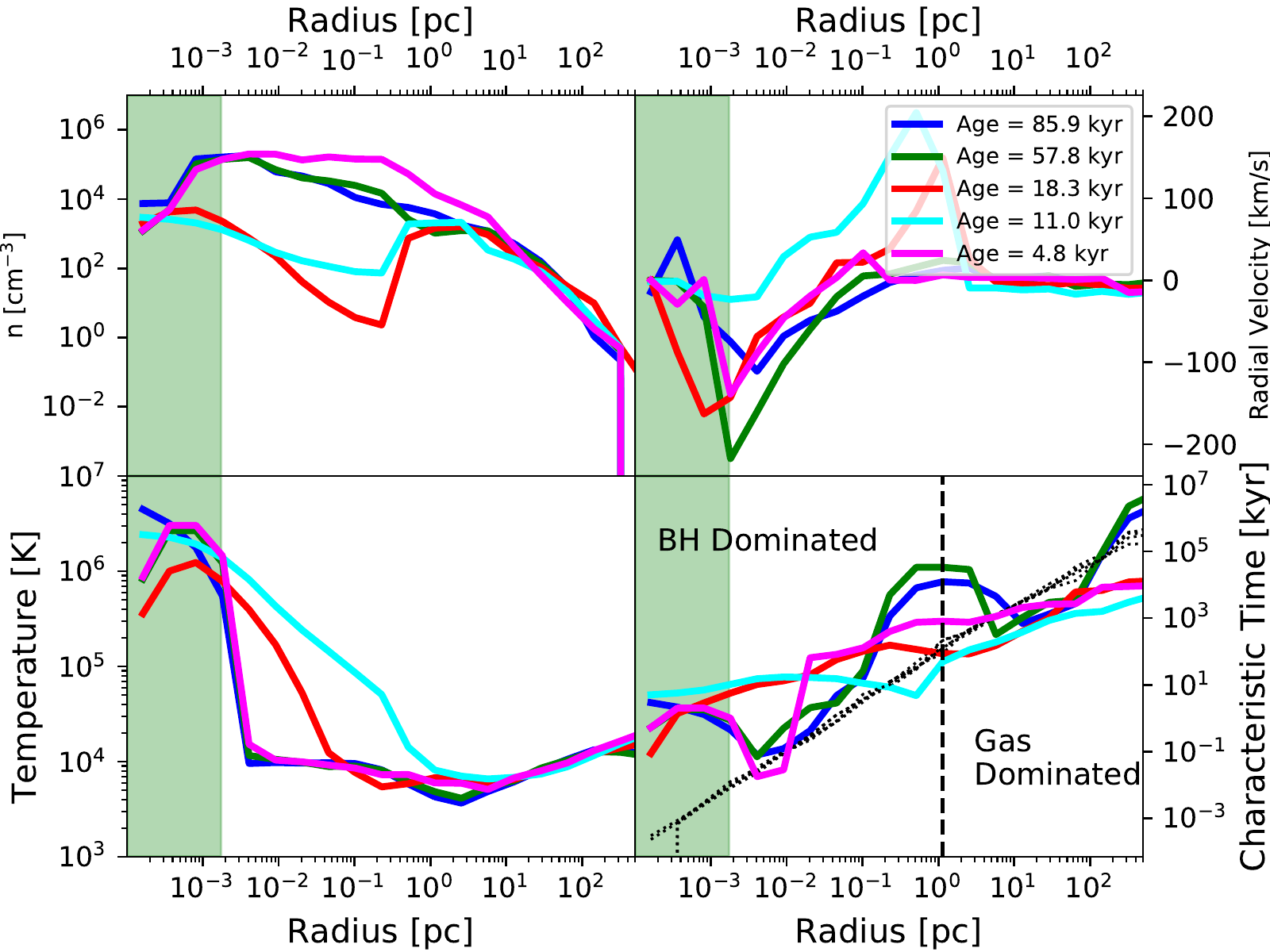}}
        \caption[]
                {\label{Radial30000}
                  The same as Figure \ref{Radial6000} for the 30,000 \kms simulation.
                  In this case we show the effect of a strong initial jet at age $\sim 11$ kyr
                  which drives gas away from the centre of the halo (see the large outflow
                  velocity at $R \sim 1$ pc). Nonetheless, the gas
                  recovers and can fall back into the centre, again reactivating accretion, within
                  a few kiloyears. 
        }
      \end{center} \end{minipage}
  \end{figure*}

%%%%%%%%%%%%%%%%%%%%%%%%%%%%%%%%%%%%%%%%%%%%%%%%%%%%%%%%%%%%%%%%%%%%%%%%%%%%
%%%%%%%%%%%%%%%%%FIGURE 6%%%%%%%%%%%%%%%%%%%%%%%%%%%%%%%%%%%%%%%%%%%%%%%
\begin{figure*}
  \centering 
  \begin{minipage}{175mm}      \begin{center}
      \centerline{
        \includegraphics[width=18cm]{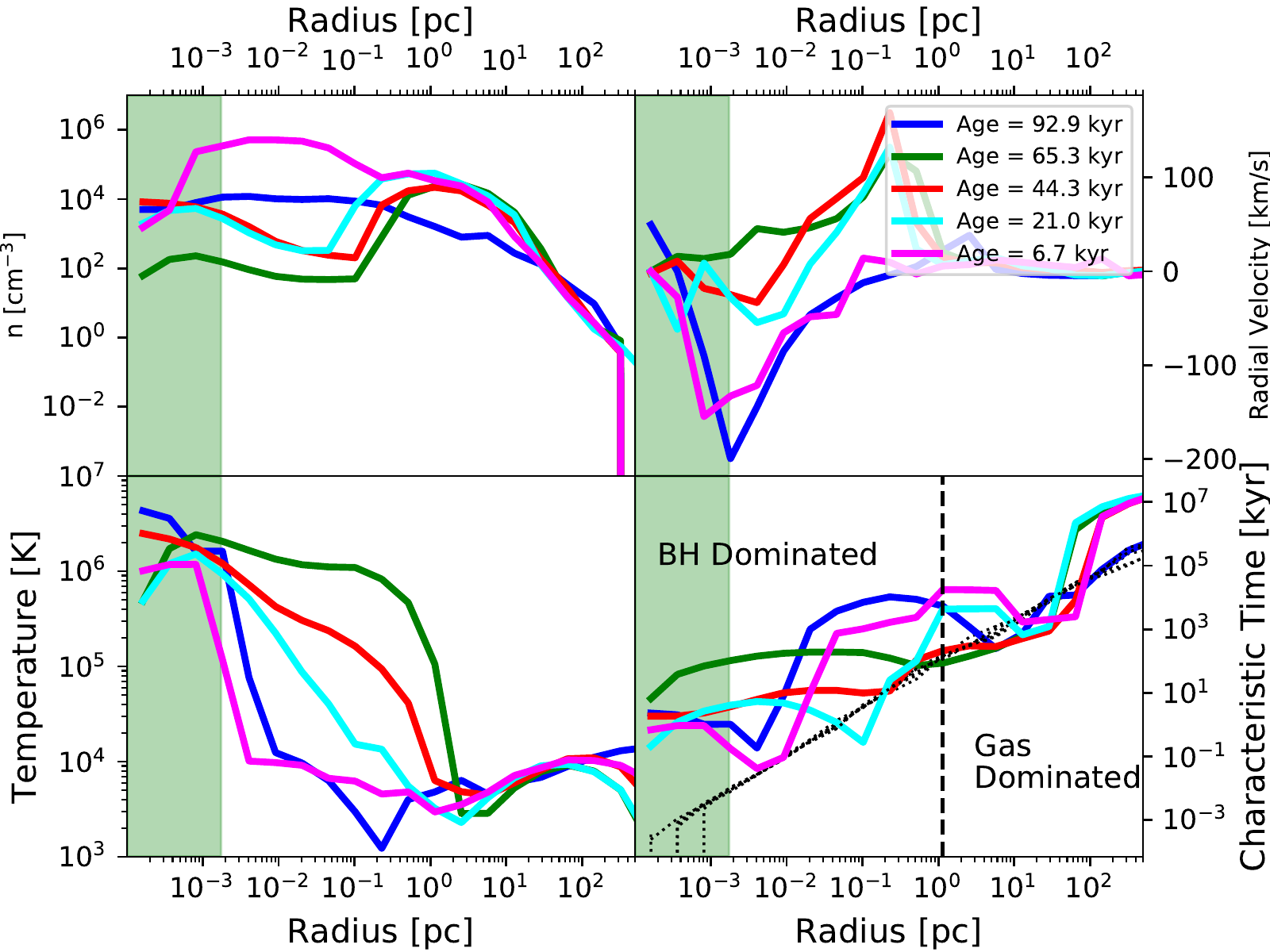}}
        \caption[]
                {\label{Radial100000}
                  The same as  Figures \ref{Radial6000} and \ref{Radial30000} for
                  the 100,000 \kms simulation. A similar pattern is observed. Jets are
                  able to effectively drive gas away from the black hole severely suppressing
                  growth. The gas recovery time can be as low as a few hundred years for the
                  highest densities. 
        }
      \end{center} \end{minipage}
  \end{figure*}

%%%%%%%%%%%%%%%%%%%%%%%%%%%%%%%%%%%%%%%%%%%%%%%%%%%%%%%%%%%%%%%%%%%%%%%%%%%%

\subsection{Simulation Realisations}
In order to test different mass loading values we select three different jet velocities. The speed of the
jet impacts the mass loading value through equation \ref{massloading}. Jets due to super-Eddington accretion
rates are launched at relativistic speeds, however, modelling relativistic jets is computationally challenging
and so mass loading the jet is often preferred. In this study we examine three different jet launching speeds:
(1) 6,000 \kms (2) 30,000 \kms and (3) 100,000 \kms. These speeds correspond to 0.018c, 0.1c and 0.33c.
As outlined in equation \ref{jetmass}, the mass ejected by the jet during each outburst is similar in all cases. Therefore, the
difference between each realisation is effectively only in the speed of the jets and hence the momentum and
energy of the jets in each realisation. As we will see, all three realisations result in similar effective
accretion rates regardless of the jet speed chosen.

%%%%%%%%%%%%%%%%%FIGURE 7%%%%%%%%%%%%%%%%%%%%%%%%%%%%%%%%%%%%%%%%%%%%%%%
\begin{figure}
  \centering 
  \includegraphics[width=0.525\textwidth]{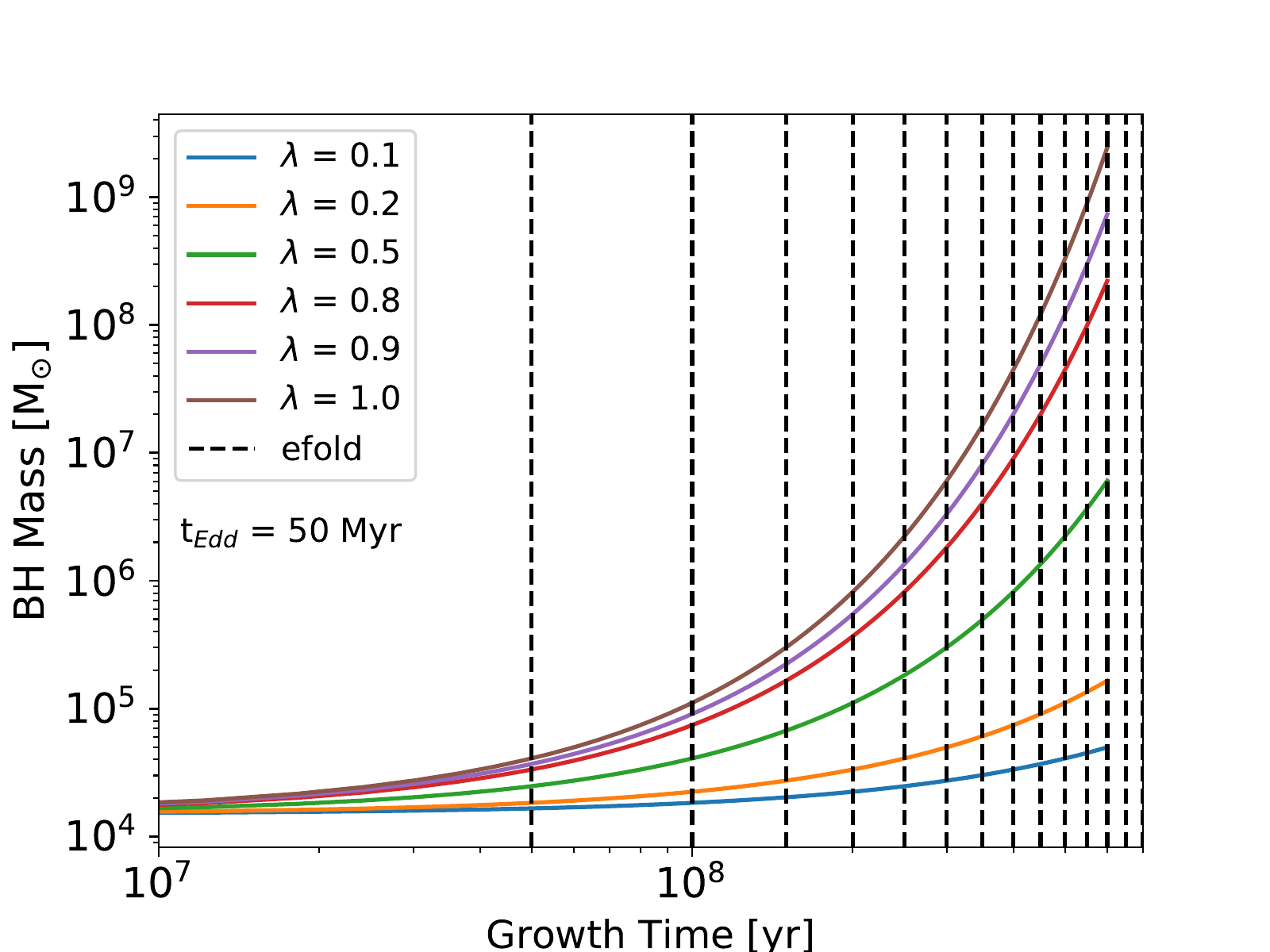}
  \caption{
    The Eddington limited growth rate of massive black holes for different
    values of the Eddington ratio, $\lambda$. The Eddington ratio is defined by
    $\lambda = \dot{M}_{BH}/\dot{M}_{Edd}$. Values of $\lambda$ which deviate even moderately
    from 1.0 have significantly reduced growth. We find that mechanical feedback
    leaded to an effective accretion rate approximately 0.25 times the canonical
    Eddington rate and hence an expected growth rate close to the orange line
    shown. The vertical dashed lines are the efolding times determined from
    the Eddington time. 
    }
    \label{EddingtonRatio}
\end{figure}

%%%%%%%%%%%%%%%%%%%%%%%%%%%%%%%%%%%%%%%%%%%%%%%%%%%%%%%%%%%%%%%%%%%%%%%%%%%%

\section{Results} \label{Sec:Results}
Our goal in this study is to examine the earliest stages of seed black hole growth immediately after the
SMS or massive PopIII collapses into a black hole. As was found in the SMS simulations in R18 the accretion rate onto
the massive PopIII star after 250 kyr is approximately 0.01 \msolarc/yr. However, feedback from an accreting black hole
is much more powerful than that of a SMS or a massive PopIII star due to the significantly enhanced compactness of the
black hole. We here investigate primarily the impact of the mechanical feedback on future accretion. \\
\indent In Figure \ref{MassAccretionRate} we plot the mass accretion rate (right panel) and the total mass
accreted by the black hole (left panel).
In the left hand panel we plot the mass accretion rate over the first 100,000 years after the seed black hole
forms for three different values of jet velocity. The x-axis runs from -50 kiloyears to 120 kiloyears.
  The negative times indicate the time for when the object was in its SMS phase. In this phase the feedback is
  modelled using radiative feedback from non-ionising radiation only, consistent with current theories of
  SMS evolution \citep{Hosokawa_2012, Regan_2018b}. At T = 0 years the SMS collapses into a black hole and
  the initial accretion rates are super-Eddington which in turn drive powerful jets. In the BH phase feedback is modelled
  using both non-ionising radiation (similar to the SMS phase) and mechanical feedback from jets. Therefore the
main difference in feedback between the negative and positive times is the mechanical feedback.\\
\indent The impact of the jets is clearly visible. Jet events are immediately
followed by periods of very low accretion before gas cools and falls back to the centre of the potential again. The
periods of high accretion, which generate the jet events, are episodic with periods of between a few kiloyears and 20
kiloyears approximately. The dynamical time of high density gas in these simulations is
  approximately 10 kiloyears at a radius of 1 pc which is consistent with the cycles of accretion
  found here. \\
\indent The green line, jets with launch velocities of 0.1 c, shows a slightly different behaviour to the two other
  realisations. In the case of the jet with velocities of 0.1 c initial periods of super-Eddington accretion are followed
  by a very large drop in accretion before it gradually rebuilds again. After the first two periods of super-Eddington
  accretion the gas never again falls in at sufficient rates to generate jets and instead the accretion remains relatively
  steady at approximately $10^{-4}$ \msolarc/yr. It is interesting that the jets with both higher and lower launch speeds show qualitatively
  different behaviour and it highlights the variability of such systems. After the gas is expelled from the
  central object it does fall back again, on approximately the dynamical time, but it need not necessarily, fall back at the same
  rates. Also this behaviour is shown to not be directly correlated with the launch speed.\\
  \indent We also plot the results of a lower resolution run (dashed red line) for the 0.33c jets. The resolution
    is reduced by a factor of 4, down from 20 levels of refinement to 18 levels of refinement. The simulations shows
    broadly the same behaviour but the detailed dynamics of the gas are different due to the lower resolution of this run. In this case the jet is spread over a larger radius with the result showing that
  accretion is suppressed for a longer period of time compared to the higher resolution case. \\
\indent In the right hand panel we plot the total mass accreted by the black hole against time. We calculate the effective
accretion rate simply by taking the initial black hole mass from the final black hole mass divided by the time. The effective
mass accretion rate is less than $10^{-4}$ \msolarc/yr in all cases. This is two orders of magnitude below the accretion rate
onto the massive PopIII star immediately prior to collapse into the black hole. Over the course of 100,000 years the black holes grow by
only approximately 10 \msolarc ~in each realisation. If accretion at this rate were to continue, the black holes would increase their mass by only one order
of magnitude over one billion years. They would grow to become intermediate mass black holes with masses of M$_{BH} \sim 10^5$
\msolar in the early Universe. By way of comparison the
    mean accretion rate during active phases is $1.3 \times 10^{-4}$ \msolarc/yr.
    We define active phases as those
    for which the black hole is accreting at more than $10^{-5}$ \msolarc/yr.\\
\indent In Figure \ref{Projection6000} we plot
projections of the number density in a 10 pc cube surrounding the black hole of the simulation of the
black hole with jet velocities of 0.018c. Similar projections of the simulations with 0.1c and 0.33c
can be found in Appendix B. Overplotted on top of the density
field is the velocity field with directional arrows. The length of the arrows is proportional to
the value of the velocity at that point. The plots are made near the start and near the end of
the simulation. Initially, in the left hand panel, we see the black hole (marked in green)
surrounded by high density gas with strong outflows due to jet events. The strong outflows are
  noticeable, at this scale,  only from the longer than average velocity line. The jets are
  highly successful at disrupting accretion onto the black hole but the impact of the jet is local.
  The jets are not able to globally influence the halo. 
  As the simulations proceed gas is driven out of the very central regions and must fall back in order
  for accretion to pick up again. The dynamical time for this system is determined by both
    the gravitational potential of the black hole, at small radii, and the self gravity of the gas
    at larger radii. We calculate the effective dynamical time as 
\begin{equation}
  t_{dyn} = \Big( {{1} \over {t_{bh}^2}} + {{1} \over {t_{ff}^2}} \big)^{-0.5}
\end{equation}
where $t_{ff}$ is the self gravity of the gas given by  $t_{ff} = \sqrt{{3} \over {32 G \rho }}$ and
$t_{bh}$ is the dynamical time of the gas within the potential of the black hole given by
$t_{bh}  = \sqrt{{R^3} \over {G*M_{BH}}}$ and $\rho$ is the gas density, $G$ is the gravitational
constant, $R$ is the radius from the black hole and $M_{BH}$ is the mass of the black hole.
For our system this corresponds to 10 kyr at a radial distance of a few parsecs
from the centre of the halo down to less than 1 kyr for the highest density gas within 0.1 pc of the
centre.
In each case what is immediately noticeable is that the outflows from the jets have little or no
effect on gas outside of approximately 1 pc. In each realisation there is no fingerprint from the
jet activity at scales larger than this even though the jets are launched with velocities of up to
100,000 \kms. The inflow from the gas is easily able to overwhelm the jet momentum, so that while
the jets are able to effectively shut off accretion in the immediate radius of the black hole
they have no effect on the gas at scales of a parsec or larger. \\
\indent In Figure \ref{FourPanel} we ``zoom-in'' to the region immediately surrounding the
  black hole in the simulation with jet velocities of 6,000 \kms. To illustrate the impact of the
  accretion events and the jet launching events we focus on visualising the behaviour of the black
  hole between 30 kyr and 40 kyr after the formation of the black hole. At this point the black hole
  is rapidly accreting material at slightly sub-Eddington accretion rates (top left and top right
  panels of Figure \ref{FourPanel}, see also Figure \ref{MassAccretionRate}). As the accretion rate
  continues to increase is eventually exceeds the Eddington rate at approximately 37 kyr resulting
  in an outflow and a decrease in the accretion rate (bottom left and bottom right panels of
  Figure \ref{FourPanel}).\\
\indent In Figures \ref{Radial6000}, \ref{Radial30000} and \ref{Radial100000} we quantify the
projection plots by taking ray profiles for different times during the course of the
  simulation. The 1D ray profiles are created by profiling the gas properties perpendicular to the
  angular momentum vector of the gas (i.e. in the plane of the accretion disk).
Focusing first on the simulation with jet velocities of 6,000 \kms (i.e. Figure \ref{Radial6000})
we see that jet events reduce the density of the gas by up to a few orders of magnitude out to a
distance of approximately 0.1 pc following at outflow. 
The low density gas that is left behind is super-heated to a temperature of $10^6$ K and the gas
is also fully ionised out to approximately 1 pc. The gas receives positive outward
  momentum from the jet events. The positive radial velocities given to the gas (as seen
  in the upper right panel as the green line) sweeps gas away from the black hole.
  This reduces the density of gas in the vicinity of the black hole and shuts off accretion.
  It is therefore primarily the momentum given to the jets that shuts off the accretion mechanics.\\
\indent Nonetheless the gas is quickly able to recover and
fall back into the centre of the potential and in a little over 10 kyr the gas has reached
sufficient density that the black
hole can accrete at very high rates and can indeed again exceed the Eddington rate
(see Figure \ref{MassAccretionRate}). This is supported by the fact that the dynamical times for the
gas between 1 pc and 10 pc is approximately 10 kyr.\\
\indent Figures \ref{Radial30000} and \ref{Radial100000} show both qualitatively and
quantitatively similar results. Periods of super-Eddington accretion launch jets, which drive
high density gas out from the centre of the halo. The density of the
gas surrounding the black hole is temporarily reduced by several orders of magnitude out to a
distance of approximately 0.1 pc. However, the gas quickly falls back to the centre of the
potential well where again gas can be accreted at high rates driving another jet event.
We note here also that the ray profiles are examining, on average, the highest density gas in
  the plane of the accretion disk. Outside of the plane of the disk the density can be much lower
  following an accretion event.

\section{Summary \& Discussion}  \label{Sec:Discussion}
In this study we examine the impact of super-Eddington accretion and feedback
on the growth rate of (supermassive) black hole seeds. In order to
create SMSs, rapid accretion onto a proto-star is required, with accretion rates
of close to 0.1 \msolarc/yr thought to be necessary to inflate the envelope
surrounding a proto-star and create a SMS. If such accretion rates can be
maintained after the collapse of the SMS then super-Eddington accretion onto the
seed black hole may be expected. \\
\indent We here investigate exactly this scenario. We use the self-consistent SMS / massive PopIII
simulations of R18 as a starting point for our simulations. The starting point is
a massive PopIII star accreting at approximately 0.01 \msolarc/yr with a final stellar mass of 
15,904 \msolarc. To examine the subsequent accretion onto a seed black hole, we artificially
collapse the massive PopIII star and create a direct collapse black hole seed.
No supernova or other feedback from the massive PopIII star is modelled to precede the formation
of a black hole. After black hole formation we model radiative feedback, below the ionisation threshold,
and mechanical feedback in the form of bipolar jets for super-Eddington accretion events. We therefore
focus almost entirely on the impact that jet feedback has on the growth of the black hole in this study.
While radiative feedback could in principle regulate accretion to the Eddington rate we do not
  model this here. We instead choose to examine the impact of super-Eddington accretion rates and the
  potential negative feedback associated with the jets driven by these extreme accretion rates. As
  discussed in \S \ref{MechanicalFeedback} we launch bipolar jets at the maximum resolution our
  setup allows. The jets are launched by spreading the velocity of the jets equally over 27 cells
  just outside the accretion radius of our black hole. This corresponds to a physical
  resolution for the jet radius of approximately 300 au. This is still much coarser than
  the radius at which jets are launched by black holes of comparable sizes and hence our
  jets may still not be sufficiently collimated. This remains an inherent limitation of these
  simulations.\\
\indent We modelled the bipolar jets using three different mass loading values. We modelled jets with velocities
of 6,000 \kms (0.018 c), 30,000 \kms (0.1c) and 100,000 \kms (0.33c). The results were qualitatively similar in
each case. Periods of super-Eddington accretion generate violent bipolar jets which suppress accretion by many
orders of magnitude. The gas surrounding the black hole is successfully evacuated due to the positive
radial velocity of the jets and the gas must wait on the free-fall time
before being available for accretion again. However, the jets are unable to break out of the
very central region of the halo. The inflow is easily able to overwhelm the outflows. We see no
impact from the jets at scales greater than approximately 1 pc and so the jets are found to be an
inherently local phenomenon with no global impact for black holes of this mass ($\sim$
15,000 \msolarc). The local impact of the jets is significant and the jets successfully shut off
the super-Eddington accretion that launched them initially, leading to periods of low accretion
immediately after jet launching, giving rise to episodic accretion as in the case of radiative
feedback from light seeds \citep{Milosavljevic_2009}. These periods of inactivity lead to
effective accretion a factor of a few below the Eddington rate. \\
\indent While a full resolution study of the results presented here is outside the scope
  of this work we did run the 0.33c simulation at 4 times lower resolution
  (see Figure \ref{MassAccretionRate}).
  We found that qualitatively the results followed the same pattern of periods of low accretion
  following a jet event followed by a return to super-Eddington accretion followed by a
  further jet event. In a future study we will explore a larger parameter space to quantitatively
  access the impact of resolution. However, tentatively our results do suggest that with increased
  resolution of the jet such a system may be able to sustain super-Eddington accretion more
  efficiently.\\
\indent The Eddington ratio, $\lambda$, can be used to describe accretion rates which are
below Eddington. In Figure \ref{EddingtonRatio} we plot the growth rate of a black hole seed
starting from the initial mass of the seed black hole studied here (M$_{\rm{init}} = 15904$ \msolarc).
The growth rate for a black hole seed is given by
\begin{equation}
  M(t) = M(t_0) \exp(t / t_{Edd})
\end{equation}
where $M(t)$ is the mass after time $t$, $M(t_0)$ is the initial seed mass at $t_0$ = 0 and $t_{Edd}$
is the Eddington (or Salpeter) time given by
\begin{equation}
  t_{Edd} =  {{\sigma_T \eta_{disk} c} \over {4 \pi G m_p}} \approx \eta_{disk} \ 5 \times 10^8 \ \ \rm{yr}
\end{equation}
with $\eta_{disk}$ the usual disk efficiency with canonical value 0.1. In this case the Eddington
time for black hole is approximately 50 Myr. The growth of the black hole is then often counted
in the number of efolding times required to reach a predetermined mass. In Figure
\ref{EddingtonRatio} we plot as dashed vertical lines the efolding times for the seed black hole
modelled here. The exponential nature of black hole growth means that initially
growth is quite sedentary and only picks up as larger efolding times are reached. The $\lambda$
factor is used to describe growth rates which are below Eddington. In this study we found that the
bipolar jets reduce growth to a factor of a few below the Eddington rate. We have also plotted these
curves in our plot assuming that both $\eta$ and $\lambda$ are time independent. Growth rates which
are a factor of two or more below Eddington have their mass, after 11 efolding times,
reduced by more than two orders of magnitude. 
\indent However, the black hole growth over longer times will likely be determined more by the
dynamics of the
host halo and its ability to merge with other haloes which may promote more efficient growth
\citep[e.g.][]{Pezzulli_2017, Valiante_2016}. If
the black hole host halo is part of a number of major mergers this will undoubtedly promote more
efficient accretion
of material. Star formation in the surrounding gas will act to diminish growth by consuming
available gas, although, we see no evidence of star formation in our simulations in the first
100,000 years after black hole formation. \\
\indent If stars start to form and gas is enriched (internally or externally), the gas distribution
will change.  On the one hand, gas will be consumed  in star formation. On the other hand, gas will
cool more easily, fragment and generate regions of high and low density.  
If the cooling gas possesses angular momentum, it will settle in a disc from which the jet can more
easily escape without damaging the surrounding 
if the jet propagates orthogonally to the disc \citep{2018MNRAS.477.1336C}. 
If none of this happens and the halo grows by accreting metal-free intergalactic gas, perhaps the
central density will be so high that the jet cannot do any damage even in its immediate
surroundings and in this case super-Eddington accretion may well be possible for extended durations.
Modelling the growth of seed black holes over several tens of megayears will require significantly
more computational power and also the identification of realistic target haloes which form a seed
in an environment which is favourable to rapid accretion at rates at or above the Eddington rate
as the black hole grows. Such simulations are likely to be possible in the near future. \\
\indent The idealised nature of our setup and the relatively short time for which we are able
to evolve our simulations mean that we are unable to provide more detailed information on the
subsequent growth of the black hole. Nonetheless, accretion onto seed black holes formed from
the direct collapse of a massive PopIII star is much more efficient that the accretion onto
PopIII remnant black holes (M$_{init} \sim $100 \msolarc)\citep[e.g.][]{Smith_2018}
which have initial accretion rates many orders of magnitude below the Eddington rate.\\
\indent As noted above we also do not model the impact of ionising radiation on the
  accretion rate of the seed black hole. The impact of ionising radiation has been modelled by
  numerous other authors \citep[e.g.][]{Milosavljevic_2009, Park_2012,  Park_2017, Sugimura_2018} in
  1D, 2D and 3D simulations. The conclusions are broadly similar - isotropic radiation feedback has
  a strongly negative impact on black hole growth. More idealised models
  \citep{Sugimura_2017, Inayoshi_2018} have shown circumstances where the impact of radiative feedback
  can be overcome but these models have not yet translated over to more general realisations.
  Either way the broad conclusions are that radiative feedback, in general, leads to a strongly
  negative impact on the accretion rate. Added to this scenario the efficient removal of angular
  momentum of gas from the system is another hurdle that must be understood
  \citep[e.g.][]{Krumholz_2005,
  Sugimura_2018}. Similar to the results found here the general conclusion is
  that feedback (be it radiative or mechanical) has a detrimental impact on black hole growth, more
  detailed investigations of the black hole environment, conditions and circumstances to
  achieve maximal growth are still required.

\section{Conclusions}
Mechanical feedback from bipolar jets is able to quickly evacuate high density gas from the
accreting black hole once the accretion rate exceeds the canonical Eddington rate. However, the gas
quickly recombines and falls back towards the centre of the potential on the freefall time of the
system. The impact of the jet outflows is local to the immediate surroundings of the black hole.
We find that the bipolar jets are unable to break out of the halo and indeed have no impact on scales
greater than approximately 1 pc. Nonetheless, the jets have a significant impact on the black hole
accretion rate.  The effective accretion rate,
taking into account periods of high accretion and intervening periods of quiescence, is reduced
by a factor of a few below the Eddington accretion rate for at least the first 100,000 years
after the formation of the black hole.\\
\indent As an example, a reduction in the black hole accretion rate of a factor of 2 below the
Eddington rate, if it were to remain at this level over the first 500 hundred million years of
the black hole growth, would reduce the mass of the black hole by a factor of
at least 20. The mass after 10 efolding times would be between $10^5$ \msolar and $10^6$ \msolar
assuming the black hole continues to accrete at half the Eddington rate.
Therefore, direct collapse black holes born into atomic cooling haloes will require external
influences (e.g. rapid major mergers with other haloes) to promote efficient accretion and
reach SMBH masses within a few hundred million years.
Further investigation of rapidly growing direct collapse host haloes will be required in the
coming years to test the growth prospects of massive black hole seeds in realistic haloes.

\section*{Acknowledgements}

J.A.R. acknowledges the support of the EU Commission through the
Marie Sk\l{}odowska-Curie Grant - ``SMARTSTARS" - grant number 699941.
M.V., A.L. and M.T. acknowledge funding from the European Research Council under the European
Community's Seventh Framework Programme (FP7/2007-2013 Grant Agreement no.\ 614199,
project ``BLACK'').  R.S.B acknowledges funding from the Centre National de la Recherche
Scientifique (CNRS) on grant ANR-16-CE31-0011. 
Computations described in this work were performed using the 
publicly-available \enzo code (http://enzo-project.org), which is the product of a collaborative 
effort of many independent scientists from numerous institutions around the world.  Their 
commitment to open science has helped make this work possible. The freely available astrophysical 
analysis code YT \citep{YT} was used to construct numerous plots within this paper. The authors 
would like to extend their gratitude to Matt Turk et al. for an excellent software package.
J.A.R. would like to thank Lydia Heck and all of the support staff involved with Durham's COSMA4
and DiRAC's COSMA5 systems for their technical support. This work was supported by the Science
and Technology Facilities Council (grant numbers ST/L00075X/1 and RF040365). This work used the
DiRAC Data Centric system at Durham University,  operated  by  the  Institute  for  Computational
Cosmology on behalf of the STFC DiRAC HPC Facility  (www.dirac.ac.uk). This equipment was funded
by BIS National E-infrastructure capital grant ST/K00042X/1, STFC capital grant ST/H008519/1,
and STFC DiRAC Operations grant ST/K003267/1 and Durham University.  DiRAC is part of the
National E-Infrastructure. 
\noindent

\bibliographystyle{mn2e_w}

\begin{thebibliography}{117}
\providecommand{\natexlab}[1]{#1}

\bibitem[{{Abel} et~al.(1997){Abel}, {Anninos}, {Zhang} \&
  {Norman}}]{Abel_1997}
{Abel} T., {Anninos} P., {Zhang} Y., {Norman} M.~L., 1997, New Astronomy, 2,
  181

\bibitem[{{Abramowicz} \& {Fragile}(2013)}]{Abramowicz_2013}
{Abramowicz} M.~A., {Fragile} P.~C., 2013, Living Reviews in Relativity, 16, 1

\bibitem[{{Abramowicz} et~al.(1988){Abramowicz}, {Czerny}, {Lasota} \&
  {Szuszkiewicz}}]{Abramowicz_1988}
{Abramowicz} M.~A., {Czerny} B., {Lasota} J.~P., {Szuszkiewicz} E., 1988, \apj,
  332, 646

\bibitem[{{Alvarez} et~al.(2009){Alvarez}, {Wise} \& {Abel}}]{Alvarez_2009}
{Alvarez} M.~A., {Wise} J.~H., {Abel} T., 2009, \apjl, 701, L133

\bibitem[{{Anninos} \& {Norman}(1994)}]{Anninos_1994}
{Anninos} W.~Y., {Norman} M.~J., 1994, \apj, 429, 434

\bibitem[{{Begelman} et~al.(2006){Begelman}, {Volonteri} \&
  {Rees}}]{Begelman_2006}
{Begelman} M.~C., {Volonteri} M., {Rees} M.~J., 2006, \mnras, 370, 289

\bibitem[{{Begelman} et~al.(2008){Begelman}, {Rossi} \&
  {Armitage}}]{Begelman_2008}
{Begelman} M.~C., {Rossi} E.~M., {Armitage} P.~J., 2008, \mnras, 387, 1649

\bibitem[{{Bondi}(1952)}]{Bondi_1952}
{Bondi} H., 1952, \mnras, 112, 195

\bibitem[{{Bromm} \& {Loeb}(2003)}]{Bromm_2003}
{Bromm} V., {Loeb} A., 2003, \apj, 596, 34

\bibitem[{{Bryan} et~al.(2014){Bryan}, {Norman}, {O'Shea}, {Abel}, {Wise},
  {Turk} \& {The Enzo Collaboration}}]{Enzo_2014}
{Bryan} G.~L., {Norman} M.~L., {O'Shea} B.~W., {Abel} T., {Wise} J.~H., {Turk}
  M.~J., {The Enzo Collaboration}, 2014, \apjs, 211, 19

\bibitem[{{Chandrasekhar}(1964)}]{Chandrasekhar_1964b}
{Chandrasekhar} S., 1964, \apj, 140, 417

\bibitem[{{Chen} et~al.(2014){Chen}, {Wise}, {Norman}, {Xu} \&
  {O'Shea}}]{Chen_2014}
{Chen} P., {Wise} J.~H., {Norman} M.~L., {Xu} H., {O'Shea} B.~W., 2014, \apj,
  795, 144

\bibitem[{{Chon} et~al.(2017){Chon}, {Hosokawa} \& {Yoshida}}]{Chon_2017b}
{Chon} S., {Hosokawa} T., {Yoshida} N., 2017, ArXiv e-prints

\bibitem[{{Cielo} et~al.(2018){Cielo}, {Bieri}, {Volonteri}, {Wagner} \&
  {Dubois}}]{2018MNRAS.477.1336C}
{Cielo} S., {Bieri} R., {Volonteri} M., {Wagner} A.~Y., {Dubois} Y., 2018,
  \mnras, 477, 1336

\bibitem[{{Ciotti} \& {Ostriker}(2001)}]{Ciotti_2001}
{Ciotti} L., {Ostriker} J.~P., 2001, \apj, 551, 131

\bibitem[{{Coppola} et~al.(2011){Coppola}, {Longo}, {Capitelli}, {Palla} \&
  {Galli}}]{Coppola_2011}
{Coppola} C.~M., {Longo} S., {Capitelli} M., {Palla} F., {Galli} D., 2011,
  \apjs, 193, 7

\bibitem[{{Coppola} et~al.(2012){Coppola}, {D'Introno}, {Galli}, {Tennyson} \&
  {Longo}}]{Coppola_2012}
{Coppola} C.~M., {D'Introno} R., {Galli} D., {Tennyson} J., {Longo} S., 2012,
  \apjs, 199, 16

\bibitem[{{Devecchi} \& {Volonteri}(2009)}]{Devecchi_2008}
{Devecchi} B., {Volonteri} M., 2009, \apj, 694, 302

\bibitem[{{Dijkstra} et~al.(2008){Dijkstra}, {Haiman}, {Mesinger} \&
  {Wyithe}}]{Dijkstra_2008}
{Dijkstra} M., {Haiman} Z., {Mesinger} A., {Wyithe} J.~S.~B., 2008, \mnras,
  391, 1961

\bibitem[{{Dijkstra} et~al.(2014){Dijkstra}, {Ferrara} \&
  {Mesinger}}]{Dijkstra_2014}
{Dijkstra} M., {Ferrara} A., {Mesinger} A., 2014, \mnras, 442, 2036

\bibitem[{{Doeleman} et~al.(2012)}]{Doeleman_2012}
{Doeleman} S.~S. et~al., 2012, Science, 338, 355

\bibitem[{{Done} et~al.(2007){Done}, {Gierli{\'n}ski} \& {Kubota}}]{Done_2007}
{Done} C., {Gierli{\'n}ski} M., {Kubota} A., 2007, \aapr, 15, 1

\bibitem[{{Done} et~al.(2012){Done}, {Davis}, {Jin}, {Blaes} \&
  {Ward}}]{Done_2012}
{Done} C., {Davis} S.~W., {Jin} C., {Blaes} O., {Ward} M., 2012, \mnras, 420,
  1848

\bibitem[{{Du} et~al.(2018)}]{Du_2018}
{Du} P. et~al., 2018, \apj, 856, 6

\bibitem[{{Dubois} et~al.(2012){Dubois}, {Pichon}, {Haehnelt}, {Kimm}, {Slyz},
  {Devriendt} \& {Pogosyan}}]{Dubois_2012}
{Dubois} Y., {Pichon} C., {Haehnelt} M., {Kimm} T., {Slyz} A., {Devriendt} J.,
  {Pogosyan} D., 2012, \mnras, 423, 3616

\bibitem[{{Eisenstein} \& {Loeb}(1995)}]{Eisenstein_1995b}
{Eisenstein} D.~J., {Loeb} A., 1995, \apj, 443, 11

\bibitem[{{Fan} et~al.(2006)}]{Fan_06}
{Fan} X. et~al., 2006, \aj, 131, 1203

\bibitem[{{Fernandez} et~al.(2014){Fernandez}, {Bryan}, {Haiman} \&
  {Li}}]{Fernandez_2014}
{Fernandez} R., {Bryan} G.~L., {Haiman} Z., {Li} M., 2014, \mnras, 439, 3798

\bibitem[{{Freitag} et~al.(2006){Freitag}, {G{\"u}rkan} \&
  {Rasio}}]{Freitag_2006}
{Freitag} M., {G{\"u}rkan} M.~A., {Rasio} F.~A., 2006, \mnras, 368, 141

\bibitem[{{Glover}(2015{\natexlab{a}})}]{Glover_2015a}
{Glover} S.~C.~O., 2015{\natexlab{a}}, \mnras, 451, 2082

\bibitem[{{Glover}(2015{\natexlab{b}})}]{Glover_2015b}
{Glover} S.~C.~O., 2015{\natexlab{b}}, \mnras, 453, 2901

\bibitem[{{Glover} \& {Abel}(2008)}]{GloverAbel_2008}
{Glover} S.~C.~O., {Abel} T., 2008, \mnras, 388, 1627

\bibitem[{{Glover} \& {Jappsen}(2007)}]{GloverJappsen_2007}
{Glover} S.~C.~O., {Jappsen} A.~K., 2007, \apj, 666, 1

\bibitem[{{Glover} \& {Savin}(2009)}]{GloverSavin_2009}
{Glover} S.~C.~O., {Savin} D.~W., 2009, \mnras, 393, 911

\bibitem[{{G{\"u}rkan} et~al.(2004){G{\"u}rkan}, {Freitag} \&
  {Rasio}}]{Gurkan_2004}
{G{\"u}rkan} M.~A., {Freitag} M., {Rasio} F.~A., 2004, \apj, 604, 632

\bibitem[{{G{\"u}rkan} et~al.(2006){G{\"u}rkan}, {Fregeau} \&
  {Rasio}}]{Gurkan_2006}
{G{\"u}rkan} M.~A., {Fregeau} J.~M., {Rasio} F.~A., 2006, \apjl, 640, L39

\bibitem[{{Habouzit} et~al.(2017){Habouzit}, {Volonteri} \&
  {Dubois}}]{Habouzit_2017}
{Habouzit} M., {Volonteri} M., {Dubois} Y., 2017, \mnras, 468, 3935

\bibitem[{{Haemmerl{\'e}} et~al.(2017){Haemmerl{\'e}}, {Woods}, {Klessen},
  {Heger} \& {Whalen}}]{Haemmerle_2017b}
{Haemmerl{\'e}} L., {Woods} T.~E., {Klessen} R.~S., {Heger} A., {Whalen} D.~J.,
  2017, ArXiv e-prints

\bibitem[{{Haemmerl{\'e}} et~al.(2018){Haemmerl{\'e}}, {Woods}, {Klessen},
  {Heger} \& {Whalen}}]{Haemmerle_2017}
{Haemmerl{\'e}} L., {Woods} T.~E., {Klessen} R.~S., {Heger} A., {Whalen} D.~J.,
  2018, \mnras, 474, 2757

\bibitem[{{Hahn} \& {Abel}(2011)}]{Hahn_2011}
{Hahn} O., {Abel} T., 2011, \mnras, 415, 2101

\bibitem[{{Heger} et~al.(2003){Heger}, {Fryer}, {Woosley}, {Langer} \&
  {Hartmann}}]{Heger_2003}
{Heger} A., {Fryer} C.~L., {Woosley} S.~E., {Langer} N., {Hartmann} D.~H.,
  2003, \apj, 591, 288

\bibitem[{{Hirano} et~al.(2017){Hirano}, {Hosokawa}, {Yoshida} \&
  {Kuiper}}]{Hirano_2017}
{Hirano} S., {Hosokawa} T., {Yoshida} N., {Kuiper} R., 2017, Science, 357, 1375

\bibitem[{{Hockney} \& {Eastwood}(1988)}]{Hockney_1988}
{Hockney} R.~W., {Eastwood} J.~W., 1988, {Computer simulation using particles}.
  Bristol: Hilger, 1988

\bibitem[{{Hosokawa} et~al.(2013{\natexlab{a}}){Hosokawa}, {Omukai} \&
  {Yorke}}]{Hosokawa_2012}
{Hosokawa} T., {Omukai} K., {Yorke} H.~W., 2013{\natexlab{a}}, \apj, 778, 178

\bibitem[{{Hosokawa} et~al.(2013{\natexlab{b}}){Hosokawa}, {Yorke}, {Inayoshi},
  {Omukai} \& {Yoshida}}]{Hosokawa_2013}
{Hosokawa} T., {Yorke} H.~W., {Inayoshi} K., {Omukai} K., {Yoshida} N.,
  2013{\natexlab{b}}, \apj, 778, 178

\bibitem[{{Hoyle} \& {Lyttleton}(1939)}]{HoyleLyttleton_1939}
{Hoyle} F., {Lyttleton} R.~A., 1939, Proceedings of the Cambridge Philosophical
  Society, 35, 405

\bibitem[{{Hoyle} \& {Lyttleton}(1940{\natexlab{a}})}]{HoyleLyttleton_1940a}
{Hoyle} F., {Lyttleton} R.~A., 1940{\natexlab{a}}, Proceedings of the Cambridge
  Philosophical Society, 36, 325

\bibitem[{{Hoyle} \& {Lyttleton}(1940{\natexlab{b}})}]{HoyleLyttleton_1940}
{Hoyle} F., {Lyttleton} R.~A., 1940{\natexlab{b}}, Proceedings of the Cambridge
  Philosophical Society, 36, 424

\bibitem[{{Inayoshi} \& {Omukai}(2012)}]{Inayoshi_2012}
{Inayoshi} K., {Omukai} K., 2012, \mnras, 422, 2539

\bibitem[{{Inayoshi} et~al.(2014){Inayoshi}, {Omukai} \&
  {Tasker}}]{Inayoshi_2014}
{Inayoshi} K., {Omukai} K., {Tasker} E., 2014, \mnras, 445, L109

\bibitem[{{Inayoshi} et~al.(2015){Inayoshi}, {Visbal} \&
  {Kashiyama}}]{Inayoshi_2015}
{Inayoshi} K., {Visbal} E., {Kashiyama} K., 2015, \mnras, 453, 1692

\bibitem[{{Inayoshi} et~al.(2016){Inayoshi}, {Haiman} \&
  {Ostriker}}]{Inayoshi_2018}
{Inayoshi} K., {Haiman} Z., {Ostriker} J.~P., 2016, \mnras, 459, 3738

\bibitem[{{Jeon} et~al.(2012){Jeon}, {Pawlik}, {Greif}, {Glover}, {Bromm},
  {Milosavljevi{\'c}} \& {Klessen}}]{Jeon_2012}
{Jeon} M., {Pawlik} A.~H., {Greif} T.~H., {Glover} S.~C.~O., {Bromm} V.,
  {Milosavljevi{\'c}} M., {Klessen} R.~S., 2012, \apj, 754, 34

\bibitem[{{Jiang} et~al.(2017){Jiang}, {Stone} \& {Davis}}]{Jiang_2017}
{Jiang} Y.~F., {Stone} J., {Davis} S.~W., 2017, ArXiv e-prints

\bibitem[{{Johnson} \& {Bromm}(2007)}]{Johnson_2007}
{Johnson} J.~L., {Bromm} V., 2007, \mnras, 374, 1557

\bibitem[{{Katz} et~al.(2015){Katz}, {Sijacki} \& {Haehnelt}}]{Katz_2015}
{Katz} H., {Sijacki} D., {Haehnelt} M.~G., 2015, \mnras, 451, 2352

\bibitem[{{Kim} et~al.(2011){Kim}, {Wise}, {Alvarez} \& {Abel}}]{Kim_2011}
{Kim} J.~h., {Wise} J.~H., {Alvarez} M.~A., {Abel} T., 2011, \apj, 738, 54

\bibitem[{{Kitsionas} \& {Whitworth}(2002)}]{Kitsionas_2002}
{Kitsionas} S., {Whitworth} A.~P., 2002, \mnras, 330, 129

\bibitem[{{Krumholz} et~al.(2004){Krumholz}, {McKee} \&
  {Klein}}]{Krumholz_2004}
{Krumholz} M.~R., {McKee} C.~F., {Klein} R.~I., 2004, \apj, 611, 399

\bibitem[{{Krumholz} et~al.(2005){Krumholz}, {McKee} \&
  {Klein}}]{Krumholz_2005}
{Krumholz} M.~R., {McKee} C.~F., {Klein} R.~I., 2005, \apj, 618, 757

\bibitem[{{Latif} et~al.(2015){Latif}, {Bovino}, {Grassi}, {Schleicher} \&
  {Spaans}}]{Latif_2015}
{Latif} M.~A., {Bovino} S., {Grassi} T., {Schleicher} D.~R.~G., {Spaans} M.,
  2015, \mnras, 446, 3163

\bibitem[{{Lupi} et~al.(2016){Lupi}, {Haardt}, {Dotti}, {Fiacconi}, {Mayer} \&
  {Madau}}]{Lupi_2016}
{Lupi} A., {Haardt} F., {Dotti} M., {Fiacconi} D., {Mayer} L., {Madau} P.,
  2016, \mnras, 456, 2993

\bibitem[{{Madau} et~al.(2014){Madau}, {Haardt} \& {Dotti}}]{Madau_2014}
{Madau} P., {Haardt} F., {Dotti} M., 2014, \apjl, 784, L38

\bibitem[{{Mayer} et~al.(2010){Mayer}, {Kazantzidis}, {Escala} \&
  {Callegari}}]{Mayer_2010}
{Mayer} L., {Kazantzidis} S., {Escala} A., {Callegari} S., 2010, \nat, 466,
  1082

\bibitem[{{Mayer} et~al.(2015){Mayer}, {Fiacconi}, {Bonoli}, {Quinn}, {Ro{\v
  s}kar}, {Shen} \& {Wadsley}}]{Mayer_2014}
{Mayer} L., {Fiacconi} D., {Bonoli} S., {Quinn} T., {Ro{\v s}kar} R., {Shen}
  S., {Wadsley} J., 2015, \apj, 810, 51

\bibitem[{{Merloni} \& {Heinz}(2008)}]{Merloni_2008}
{Merloni} A., {Heinz} S., 2008, ArXiv e-prints, 805

\bibitem[{{Milosavljevi{\'c}} et~al.(2009){Milosavljevi{\'c}}, {Couch} \&
  {Bromm}}]{Milosavljevic_2009}
{Milosavljevi{\'c}} M., {Couch} S.~M., {Bromm} V., 2009, \apjl, 696, L146

\bibitem[{{Mitsuda} et~al.(1984)}]{Mitsuda_1984}
{Mitsuda} K. et~al., 1984, \pasj, 36, 741

\bibitem[{{Mortlock} et~al.(2011)}]{Mortlock_11}
{Mortlock} D.~J. et~al., 2011, \nat, 474, 616

\bibitem[{{Omukai} et~al.(2008){Omukai}, {Schneider} \& {Haiman}}]{Omukai_2008}
{Omukai} K., {Schneider} R., {Haiman} Z., 2008, \apj, 686, 801

\bibitem[{{O'Shea} et~al.(2005){O'Shea}, {Abel}, {Whalen} \&
  {Norman}}]{OShea_2005b}
{O'Shea} B.~W., {Abel} T., {Whalen} D., {Norman} M.~L., 2005, \apj, 628, L5

\bibitem[{{O'Shea} et~al.(2015){O'Shea}, {Wise}, {Xu} \& {Norman}}]{OShea_2015}
{O'Shea} B.~W., {Wise} J.~H., {Xu} H., {Norman} M.~L., 2015, \apjl, 807, L12

\bibitem[{{Pacucci} et~al.(2015){Pacucci}, {Volonteri} \&
  {Ferrara}}]{Pacucci_2015a}
{Pacucci} F., {Volonteri} M., {Ferrara} A., 2015, \mnras, 452, 1922

\bibitem[{{Park} \& {Ricotti}(2012)}]{Park_2012}
{Park} K., {Ricotti} M., 2012, \apj, 747, 9

\bibitem[{{Park} et~al.(2017){Park}, {Wise} \& {Bogdanovi{\'c}}}]{Park_2017}
{Park} K., {Wise} J.~H., {Bogdanovi{\'c}} T., 2017, \apj, 847, 70

\bibitem[{{Pezzulli} et~al.(2016){Pezzulli}, {Valiante} \&
  {Schneider}}]{Pezzulli_2016}
{Pezzulli} E., {Valiante} R., {Schneider} R., 2016, \mnras, 458, 3047

\bibitem[{{Pezzulli} et~al.(2017){Pezzulli}, {Volonteri}, {Schneider} \&
  {Valiante}}]{Pezzulli_2017}
{Pezzulli} E., {Volonteri} M., {Schneider} R., {Valiante} R., 2017, \mnras,
  471, 589

\bibitem[{{Portegies Zwart} et~al.(2004){Portegies Zwart}, {Baumgardt}, {Hut},
  {Makino} \& {McMillan}}]{PortegiesZwart_2004}
{Portegies Zwart} S.~F., {Baumgardt} H., {Hut} P., {Makino} J., {McMillan}
  S.~L.~W., 2004, \nat, 428, 724

\bibitem[{{Regan} \& {Downes}(2018{\natexlab{a}})}]{Regan_2018a}
{Regan} J.~A., {Downes} T.~P., 2018{\natexlab{a}}, \mnras, 475, 4636

\bibitem[{{Regan} \& {Downes}(2018{\natexlab{b}})}]{Regan_2018b}
{Regan} J.~A., {Downes} T.~P., 2018{\natexlab{b}}, \mnras, 478, 5037

\bibitem[{{Regan} \& {Haehnelt}(2009{\natexlab{a}})}]{Regan_2009b}
{Regan} J.~A., {Haehnelt} M.~G., 2009{\natexlab{a}}, \mnras, 396, 343

\bibitem[{{Regan} \& {Haehnelt}(2009{\natexlab{b}})}]{Regan_2009}
{Regan} J.~A., {Haehnelt} M.~G., 2009{\natexlab{b}}, \mnras, 393, 858

\bibitem[{{Regan} et~al.(2015){Regan}, {Johansson} \& {Wise}}]{Regan_2015}
{Regan} J.~A., {Johansson} P.~H., {Wise} J.~H., 2015, \mnras, 449, 3766

\bibitem[{{Regan} et~al.(2017){Regan}, {Visbal}, {Wise}, , {Haiman},
  {Johansson} \& {Bryan}}]{Regan_2017}
{Regan} J.~A., {Visbal} E., {Wise} J.~H., , {Haiman} Z., {Johansson} P.~H.,
  {Bryan} G.~L., 2017, Nature Astronomy, 1, 0075

\bibitem[{{Ruffert}(1994)}]{Ruffert_1994}
{Ruffert} M., 1994, \apj, 427, 342

\bibitem[{{Ruffert} \& {Arnett}(1994)}]{Ruffert_1994a}
{Ruffert} M., {Arnett} D., 1994, \apj, 427, 351

\bibitem[{{Sakurai} et~al.(2016{\natexlab{a}}){Sakurai}, {Inayoshi} \&
  {Haiman}}]{Sakurai_2016a}
{Sakurai} Y., {Inayoshi} K., {Haiman} Z., 2016{\natexlab{a}}, \mnras, 461, 4496

\bibitem[{{Sakurai} et~al.(2016{\natexlab{b}}){Sakurai}, {Vorobyov},
  {Hosokawa}, {Yoshida}, {Omukai} \& {Yorke}}]{Sakurai_2016}
{Sakurai} Y., {Vorobyov} E.~I., {Hosokawa} T., {Yoshida} N., {Omukai} K.,
  {Yorke} H.~W., 2016{\natexlab{b}}, \mnras, 459, 1137

\bibitem[{{S{\c a}dowski}(2009)}]{Sadowski_2009}
{S{\c a}dowski} A., 2009, \apjs, 183, 171

\bibitem[{{S{\c a}dowski} \& {Narayan}(2016)}]{Sadowski_2016}
{S{\c a}dowski} A., {Narayan} R., 2016, \mnras, 456, 3929

\bibitem[{{S{\c a}dowski} et~al.(2014){S{\c a}dowski}, {Narayan}, {McKinney} \&
  {Tchekhovskoy}}]{Sadowski_2014}
{S{\c a}dowski} A., {Narayan} R., {McKinney} J.~C., {Tchekhovskoy} A., 2014,
  \mnras, 439, 503

\bibitem[{{S{\c a}dowski} et~al.(2016){S{\c a}dowski}, {Lasota}, {Abramowicz}
  \& {Narayan}}]{Sadowski_2016a}
{S{\c a}dowski} A., {Lasota} J.~P., {Abramowicz} M.~A., {Narayan} R., 2016,
  \mnras, 456, 3915

\bibitem[{{Schauer} et~al.(2017){Schauer}, {Regan}, {Glover} \&
  {Klessen}}]{Schauer_2017}
{Schauer} A.~T.~P., {Regan} J., {Glover} S.~C.~O., {Klessen} R.~S., 2017,
  \mnras, 471, 4878

\bibitem[{{Schleicher} et~al.(2013){Schleicher}, {Palla}, {Ferrara}, {Galli} \&
  {Latif}}]{Schleicher_2013}
{Schleicher} D.~R.~G., {Palla} F., {Ferrara} A., {Galli} D., {Latif} M., 2013,
  \aap, 558, A59

\bibitem[{{Shu}(1977)}]{Shu_1977}
{Shu} F.~H., 1977, \apj, 214, 488

\bibitem[{{Smith} et~al.(2018){Smith}, {Regan}, {Downes}, {Norman}, {O'Shea} \&
  {Wise}}]{Smith_2018}
{Smith} B.~D., {Regan} J.~A., {Downes} T.~P., {Norman} M.~L., {O'Shea} B.~W.,
  {Wise} J.~H., 2018, \mnras, 480, 3762

\bibitem[{{Smith} et~al.(2017)}]{Grackle}
{Smith} B.~D. et~al., 2017, \mnras, 466, 2217

\bibitem[{{Stone} \& {Norman}(1992{\natexlab{a}})}]{Stone_1992}
{Stone} J.~M., {Norman} M.~L., 1992{\natexlab{a}}, \apjs, 80, 753

\bibitem[{{Stone} \& {Norman}(1992{\natexlab{b}})}]{Stone_1992b}
{Stone} J.~M., {Norman} M.~L., 1992{\natexlab{b}}, \apjs, 80, 791

\bibitem[{{Sugimura} et~al.(2014){Sugimura}, {Omukai} \&
  {Inoue}}]{Sugimura_2014}
{Sugimura} K., {Omukai} K., {Inoue} A.~K., 2014, \mnras, 445, 544

\bibitem[{{Sugimura} et~al.(2017){Sugimura}, {Hosokawa}, {Yajima} \&
  {Omukai}}]{Sugimura_2017}
{Sugimura} K., {Hosokawa} T., {Yajima} H., {Omukai} K., 2017, \mnras, 469, 62

\bibitem[{{Sugimura} et~al.(2018){Sugimura}, {Hosokawa}, {Yajima}, {Inayoshi}
  \& {Omukai}}]{Sugimura_2018}
{Sugimura} K., {Hosokawa} T., {Yajima} H., {Inayoshi} K., {Omukai} K., 2018,
  \mnras, 478, 3961

\bibitem[{{Tanaka} \& {Li}(2014)}]{Tanaka_2014}
{Tanaka} T.~L., {Li} M., 2014, \mnras, 439, 1092

\bibitem[{{Tang} et~al.(2019)}]{Tang_2019}
{Tang} J.~J. et~al., 2019, arXiv e-prints, arXiv:1901.02615

\bibitem[{{Toyouchi} et~al.(2018){Toyouchi}, {Hosokawa}, {Sugimura}, {Nakatani}
  \& {Kuiper}}]{Toyouchi_2018}
{Toyouchi} D., {Hosokawa} T., {Sugimura} K., {Nakatani} R., {Kuiper} R., 2018,
  ArXiv e-prints, arXiv:1811.01368

\bibitem[{{Tseliakhovich} \& {Hirata}(2010)}]{Tseliakhovich_2010}
{Tseliakhovich} D., {Hirata} C., 2010, \prd, 82, 083520

\bibitem[{{Turk} et~al.(2011){Turk}, {Smith}, {Oishi}, {Skory}, {Skillman},
  {Abel} \& {Norman}}]{YT}
{Turk} M.~J., {Smith} B.~D., {Oishi} J.~S., {Skory} S., {Skillman} S.~W.,
  {Abel} T., {Norman} M.~L., 2011, \apjs, 192, 9

\bibitem[{{Umeda} et~al.(2016){Umeda}, {Hosokawa}, {Omukai} \&
  {Yoshida}}]{Umeda_2016}
{Umeda} H., {Hosokawa} T., {Omukai} K., {Yoshida} N., 2016, \apj, 830, L34

\bibitem[{{Valiante} et~al.(2016){Valiante}, {Schneider}, {Volonteri} \&
  {Omukai}}]{Valiante_2016}
{Valiante} R., {Schneider} R., {Volonteri} M., {Omukai} K., 2016, \mnras, 457,
  3356

\bibitem[{{Visbal} et~al.(2014){Visbal}, {Haiman} \& {Bryan}}]{Visbal_2014b}
{Visbal} E., {Haiman} Z., {Bryan} G.~L., 2014, \mnras, 445, 1056

\bibitem[{{Wang} et~al.(2006){Wang}, {Chen} \& {Hu}}]{Wang_2006}
{Wang} J.~M., {Chen} Y.~M., {Hu} C., 2006, \apj, 637, L85

\bibitem[{{Whalen} et~al.(2004){Whalen}, {Abel} \& {Norman}}]{Whalen_2004}
{Whalen} D., {Abel} T., {Norman} M.~L., 2004, \apj, 610, 14

\bibitem[{{Wise} \& {Abel}(2011)}]{WiseAbel_2011}
{Wise} J.~H., {Abel} T., 2011, \mnras, 414, 3458

\bibitem[{{Wolcott-Green} \& {Haiman}(2012)}]{WolcottGreen_2012}
{Wolcott-Green} J., {Haiman} Z., 2012, \mnras, 425, L51

\bibitem[{{Woods} et~al.(2017){Woods}, {Heger}, {Whalen}, {Haemmerl{\'e}} \&
  {Klessen}}]{Woods_2017}
{Woods} T.~E., {Heger} A., {Whalen} D.~J., {Haemmerl{\'e}} L., {Klessen} R.~S.,
  2017, \apjl, 842, L6

\bibitem[{{Woosley} et~al.(2002){Woosley}, {Heger} \& {Weaver}}]{Woosley_2002}
{Woosley} S.~E., {Heger} A., {Weaver} T.~A., 2002, Reviews of Modern Physics,
  74, 1015

\bibitem[{{Xu} et~al.(2013){Xu}, {Wise} \& {Norman}}]{Xu_2013}
{Xu} H., {Wise} J.~H., {Norman} M.~L., 2013, \apj, 773, 83

\end{thebibliography}

\bsp	% typesetting comment

\appendix
\section{Black Hole Spectral Energy Distribution}
In this paper we model radiation from the black holes below the ionisation threshold of
hydrogen only. We omit the impact of ionising feedback so as to concentrate solely on the
mechanical feedback from the jets. Nonetheless our implementation is setup to calculate the
energy spectrum from the black hole accretion disk and we elucidate that methodology here
for the interested reader.\\
\indent The calculation is based on the assumption of a multi-colour blackbody disk surrounded
by a hot corona. The implementation within \enzo is based on a lookup table which tabulates the
SED based on the mass of the black hole and the black hole accretion rate. The masses range from
1 \msolar up to $10^{9}$ \msolar with mass accretion rates running from $10^{-6}$ \msolar/yr up to
$1^{3}$ \msolar/yr. We begin by examining the multi-color disk (MCD) component
\citep{Mitsuda_1984, Done_2007}.
The model assumes that the local emission from the disk is Planckian with a temperature profile
$T(r) \propto r^{-3/4}$. The flux from the black hole can then be written as:
\begin{equation}
  F_{MCD} = \int^{R_{inner}}_{R_{outer}} 2 \pi R B(E, T) dR
\end{equation}
where $F_{MCD}$ is the flux eminating from the MCD, $R_{inner}$ and $R_{outer}$ are the inner and
outer boundaries of the accretion disk, $B(E,T)$ is the
Planck function and $T$ is the temperature. $R_{inner}$ is set to be equal to the
inner most stable circular orbit (i.e. $R_{inner} = R_{isco} = 6 R_{sh} = 6 G M/c^2$)
where $R_{sh}$ is the Schwartzchild radius ($GM/c^2$). 
$R_{outer}$ is set to be 1000 times
$R_{sh}$. The inner disk temperature, $T_{inner}$ is given by
\begin{equation}
  T_{inner} = \Big({{3 G  M_{BH} \dot{M}_{BH}} \over {8 \pi R_{inner}^3 \sigma_{SB}}} \Big)^{0.25}
\end{equation}
where $M_{BH}$ is the mass of the black hole,  $\dot{M}_{BH}$ is the accretion rate onto the black holes
and $\sigma_{SB}$ is the Stefan-Boltzmann constant. The disk temperature, $T(R)$, is
then found by applying the scaling relation $T(r) \propto r^{-3/4}$. Once the flux for the MCD is found
what remains is to add the contribution from the corona surrounding the accretion disk. \\
\indent For modelling the contribution from the corona we apply a power law with spectral power index
of $\Gamma = -1.7$. The black hole normalisation in this case can be written as:
\begin{equation}
  BH_{norm} = {{(1 + \Gamma) L_{BH}} \over E_{end}^{\Gamma + 1} - E_{start}^{\Gamma + 1}}
\end{equation}
where $L_{BH} = \eta \dot{M}_{BH} c^2$, $E_{start}$, $E_{end}$ are the limits of the energy over
which the corona applies and $\eta$ is the radiative efficiency. $E_{start} = 200$ eV, $E_{end} = 10000$ eV. The spectral energy component
from the corona, $F_C$, can then be written as
\begin{equation}
  F_{C} = BH_{norm} E^{\Gamma}
\end{equation}
where E is the energy range of the (high-energy) corona. The total energy contribution from
  the black hole is split equally between the MCD and corona and each component is multiplied
by 0.5 in the actual calculation. \\
%%%%%%%%%%%%%%%%%Appendix FIGURE 1%%%%%%%%%%%%%%%%%%%%%%%%%%%%%%%%%%%%%%%%%%%%%%%
\begin{figure}
  \centering 
  \includegraphics[width=0.47\textwidth]{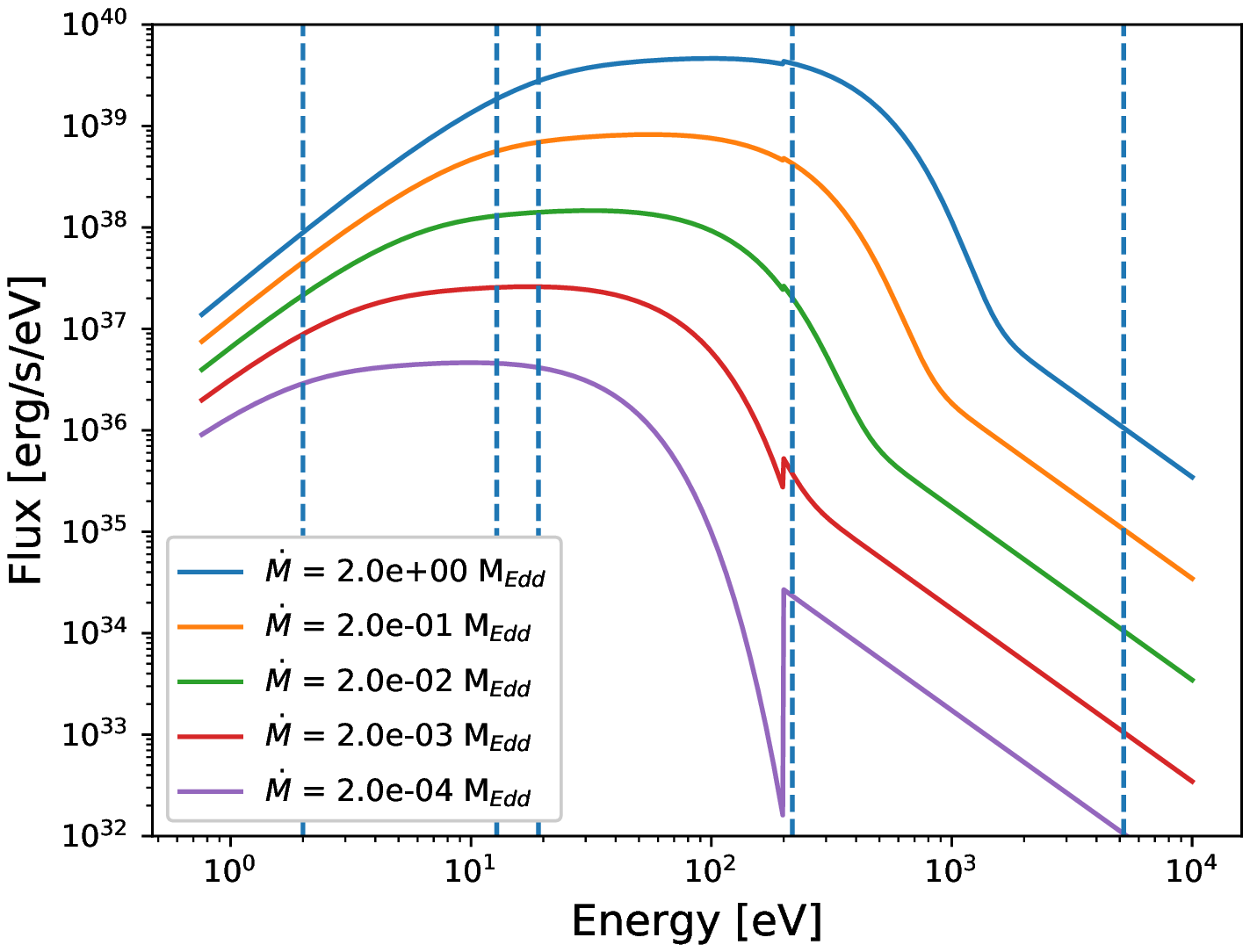}
    \caption{
      An example of the SED for the black holes used in this study. The mass of the black hole in
      this case is $M_{BH} = 15000$ \msolarc. The SED changes with both the black hole mass and the
      accretion rate. We discretise the radiation into a five bins (2.0 eV, 12.8 eV, 19.1eV,
      217.3 eV and 5190 eV). The bins are marked with dashed lines in the plot. The SED is composed
      of a multicolour disk and a power law for the high energy hard X-ray part of the spectrum.
      The contribution from the high-energy component of the spectrum falls off significantly as the
      accretion rate onto the black holes decreases. 
    }
    \label{BHSpectra}
\end{figure}

%%%%%%%%%%%%%%%%%%%%%%%%%%%%%%%%%%%%%%%%%%%%%%%%%%%%%%%%%%%%%%%%%%%%%%%%%%%%
\indent In Figure \ref{BHSpectra} we plot the SED for a selection of black hole accretion rates
for a black hole with a mass of 15,000 \msolarc. A 15,000 \msolar black hole will
experiencing super-Eddington accretion will have a peak in 
the SED of approximately 100 eV and sustained emission into the hard X-ray. As the
accretion rate deteriorates (for example after an outflow) the peak in the SED regresses to
lower energies with most of the energy lying below the ionisation threshold for hydrogen
once the accretion rate drops below approximately $10^{-3} \dot{M}_{Edd}$.

\section{Visualisations of the simulations with jets of velocity 0.1c and 0.33c.}
\begin{figure*}
  \centering 
  \begin{minipage}{175mm}      \begin{center}
      \centerline{
        \includegraphics[width=9cm]{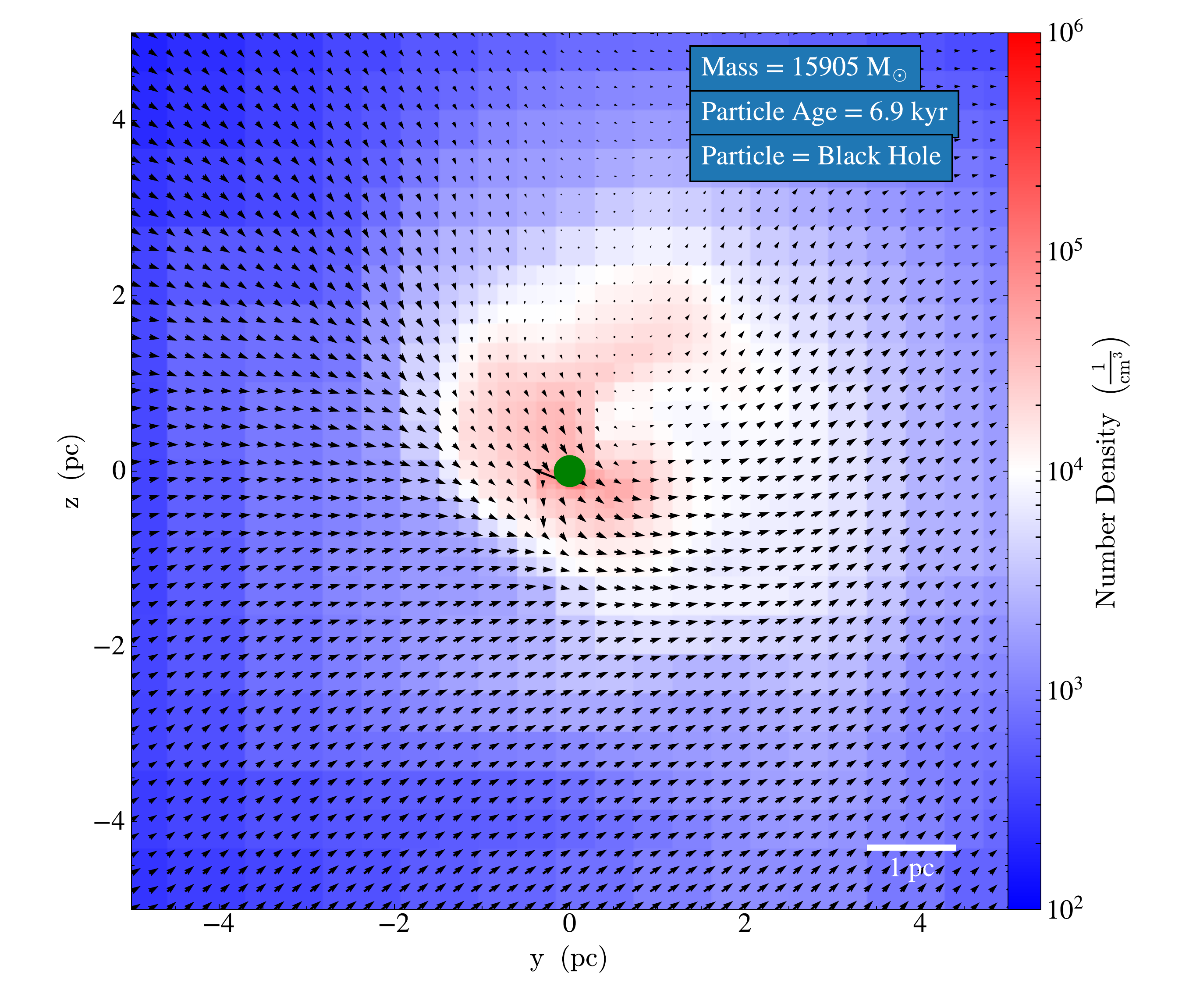}
        \includegraphics[width=9cm]{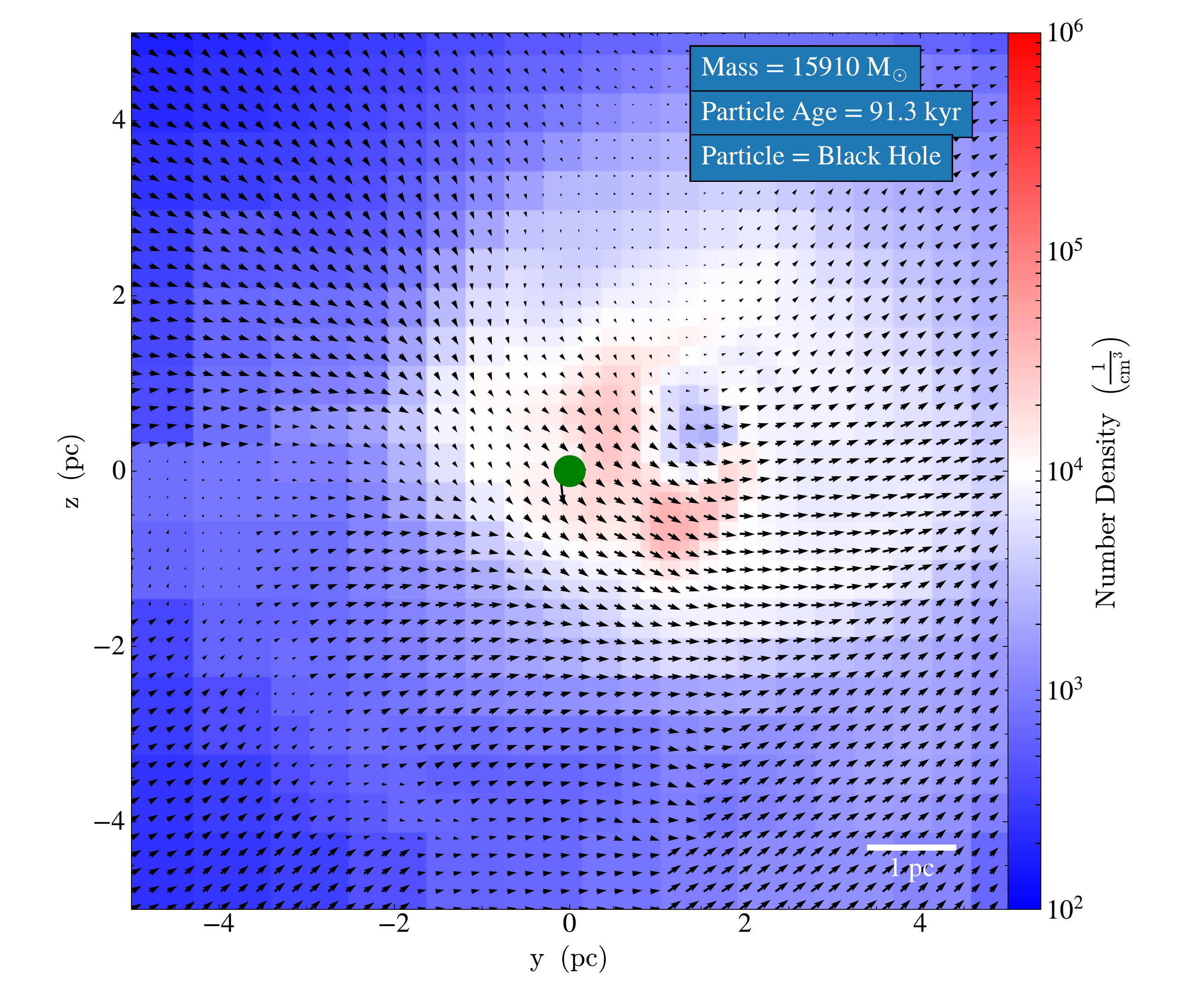}}
        \caption[]
        {\label{Projection30000}
          The same as Figure \ref{Projection6000} except for the simulation with
          jet velocities of 30,000 \kms. This simulations suffers an initial
            large drop in accretion rates after the first few jets are launched. The accretion
            rates then settles down to a more or less constant rate of $\dot{M}_{BH} \sim 10^{-4}$
            \msolarc/yr. 
        }
      \end{center} \end{minipage}
  \end{figure*}

\begin{figure*}
  \centering 
  \begin{minipage}{175mm}      \begin{center}
      \centerline{
        \includegraphics[width=9cm]{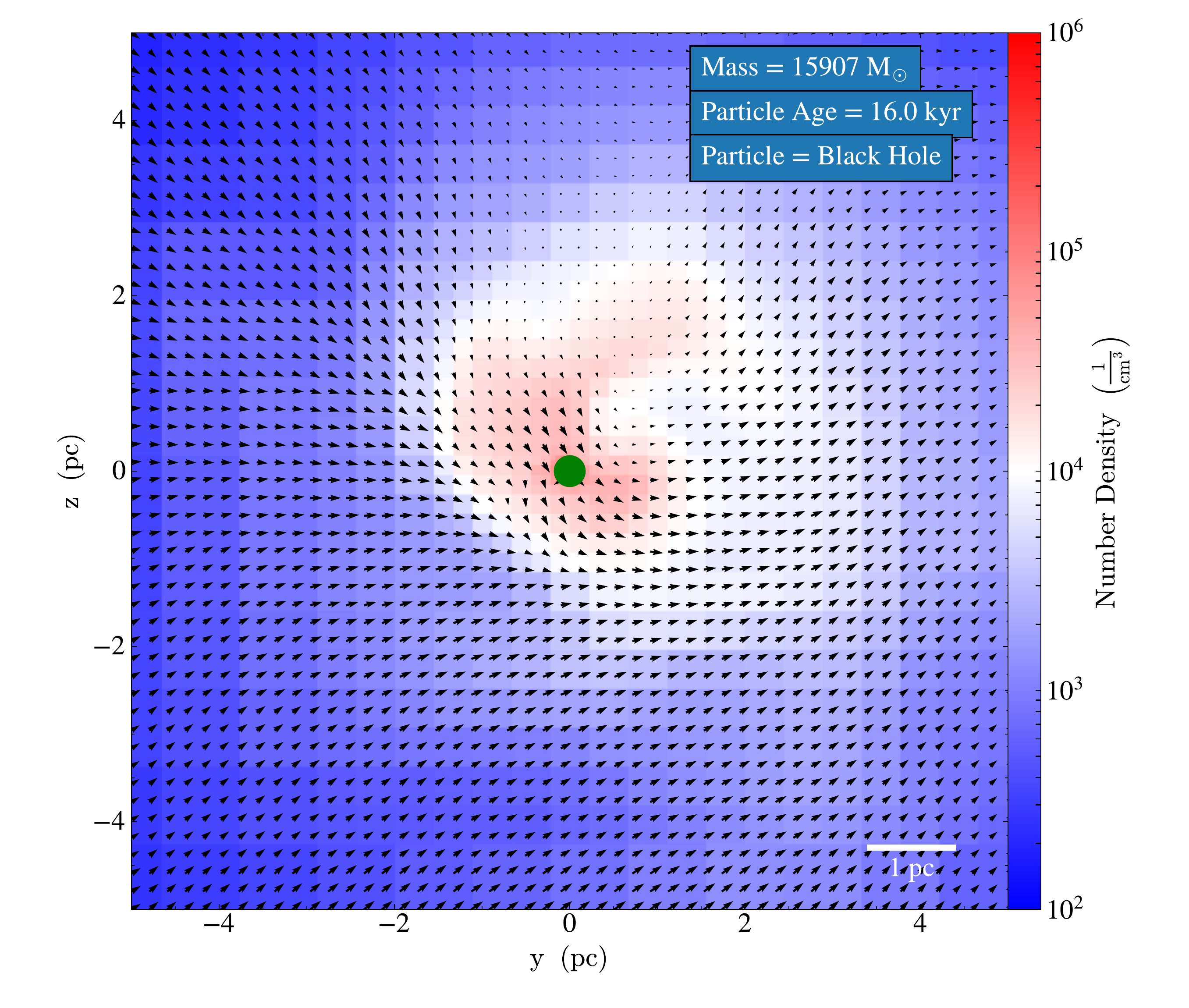}
        \includegraphics[width=9cm]{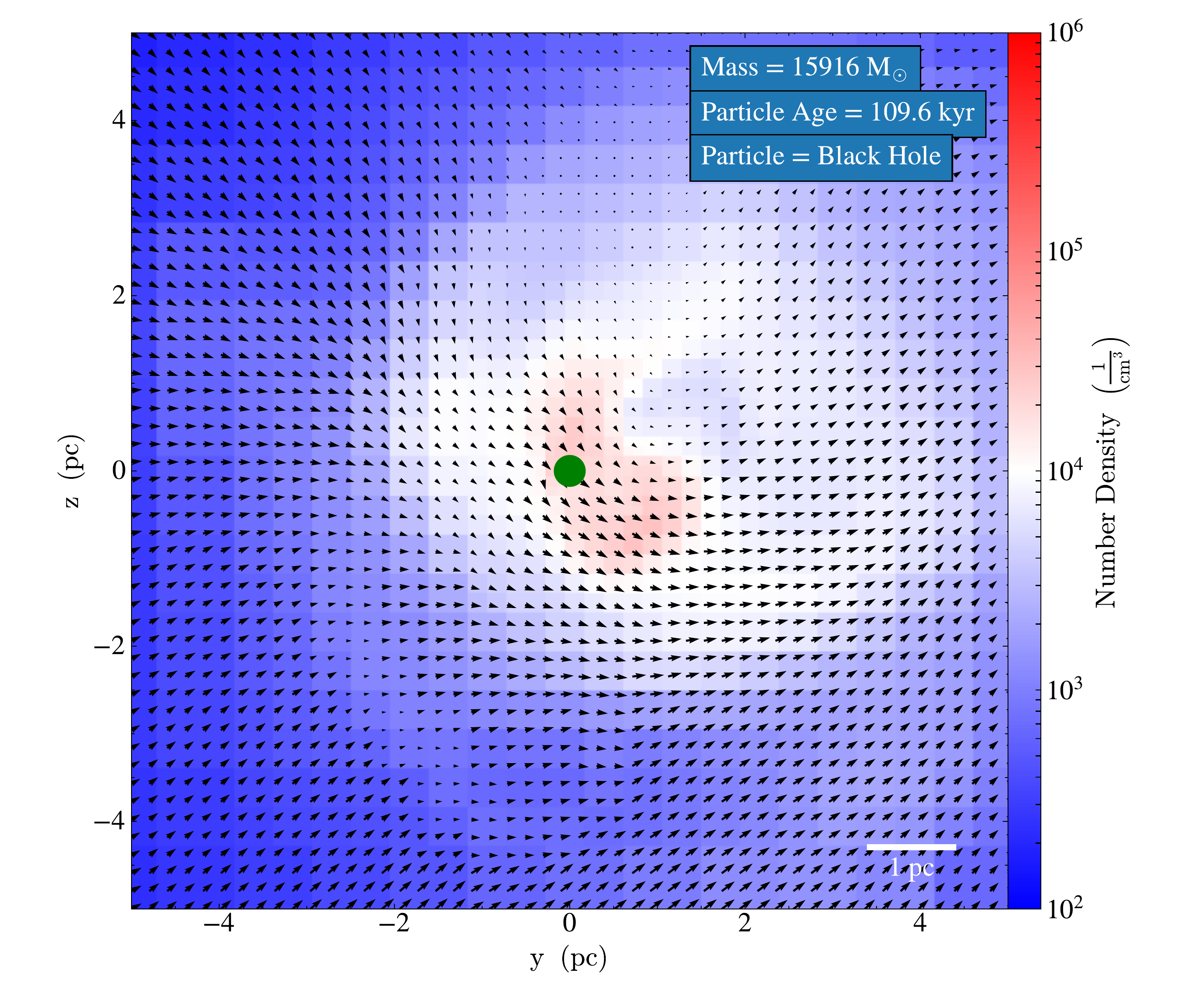}}
        \caption[]
        {\label{Projection100000}
          The same as Figures \ref{Projection6000} and \ref{Projection30000} except for the
          simulation with jet velocities of 100,000 \kms. This is the realisation with the
            highest
          jet velocities and hence the lowest mass loading. Nonetheless, even with these extremely
          high outflow velocities these is little, if any, impact from the jets at the parsec scale.
          The impact of the jets is only seen at sub-parsec scales but they can significantly hinder
          accretion. 
        }
      \end{center} \end{minipage}
  \end{figure*}

The projections in a 10 parsec cube surrounding the black hole are shown in Figures \ref{Projection30000} and \ref{Projection100000}. Qualitatively they are similar to Figure \ref{Projection6000}. 

%%%%%%%%%%%%%%%%%Appendix FIGURE 2%%%%%%%%%%%%%%%%%%%%%%%%%%%%%%%%%%%%%%%%%%%%%%%

%%%%%%%%%%%%%%%%%%%%%%%%%%%%%%%%%%%%%%%%%%%%%%%%%%%%%%%%%%%%%%%%%%%%%%%%%%%%

 %%%%%%%%%%%%%%%%%Appendix FIGURE 3%%%%%%%%%%%%%%%%%%%%%%%%%%%%%%%%%%%%%%%%%%%%%%%

%%%%%%%%%%%%%%%%%%%%%%%%%%%%%%%%%%%%%%%%%%%%%%%%%%%%%%%%%%%%%%%%%%%%%%%%%%%%
\label{lastpage}
\end{document}